\documentclass[preprint,12pt]{elsarticle}
\usepackage{xcolor}
\usepackage[utf8]{inputenc}
\usepackage{textgreek}
\usepackage[hidelinks]{hyperref}
\hypersetup{
    pdfauthor={Mohammad Parsa Rezaei, Grzegorz Kudra, Krzysztof Witkowski, Grzegorz Wasilewski,Jan Awrejcewicz},
    pdftitle={Dynamic Behavior of Shear-Thickening Fluids under Harmonic Excitation: An Experimental Investigation}
}

\usepackage[figurename=Fig.]{caption}
\usepackage{svg}
\usepackage{caption}
\usepackage{float}

\usepackage{lineno}

\usepackage{amssymb}
\usepackage{amsmath}

\journal{Mechanical Systems and Signal Processing}

\begin{document}

\begin{frontmatter}



\title{Dynamic Behavior of Shear-Thickening Fluids under Harmonic Excitation: An Experimental Investigation}



\author[a]{Mohammad Parsa Rezaei}
\author[a]{Grzegorz Kudra\corref{cor1}\fnref{label2}}
\ead{grzegorz.kudra@p.lodz.pl}
\author[a]{Krzysztof Witkowski}
\author[a]{Grzegorz Wasilewski}
\author[a]{Jan Awrejcewicz}

\affiliation[a]{%
  organization={Department of Automation, Biomechanics and Mechatronics, Lodz University of Technology},%
  addressline={Stefanowski St, 1/15},%
  city={Lodz},%
  postcode={90-924},%
  state={Lodz},%
  country={Poland}%
}

\begin{abstract}
\sloppy
 Shear-thickening fluids (STFs) become more viscous under shear stress, which makes them useful for many engineering and scientific applications. However, their behavior under normal forces, especially when these forces are applied harmonically, is less understood. Here, we examine the dynamic response of STFs under harmonic excitation. Our experimental setup features an unbalanced rotor attached to a vibrating plate submerged in an STF-filled container. We monitored the rotor’s speed, the system’s displacement, and the STF force. Data from experiments without STF were used to identify system parameters, and measurements of the STF force revealed the dynamic nature of the STF force. By comparing responses with and without STF, we identified three regions of behavior: in the pre-resonance region, the STF force is negligible; at resonance, it acts as a significant damping force; and in the post-resonance region, it behaves like an on-off force. Overall, the STF effectively reduces resonance amplitudes. These results can inform the design of complex structures incorporating STFs. 
 \end{abstract}

\begin{graphicalabstract}
\end{graphicalabstract}

\begin{highlights}
\item Distinct STF behaviors under harmonic excitation: negligible force pre-resonance, strong damping at resonance, and impact-like patterns post-resonance.
\item Oscillator model validated through parametric identification and numerical simulations, matching experimental data.
\item STF integration considerably reduces resonance amplitudes, supporting adaptive vibration control.
\end{highlights}

\begin{keyword}
Shear-Thickening Fluids, Harmonic Excitation, Parametric Identification, Dynamic Behavior, Resonance, Damping, Experimental Investigation.
\end{keyword}
\end{frontmatter}

\section{Introduction} \label{Introduction}

Shear-thickening fluids (STFs) are a class of non-Newtonian materials that exhibit a reversible and dramatic increase in viscosity when subjected to applied shear stress \cite{Hoffman1972}. This behavior enables STFs to provide tunable mechanical responses \cite{ZAREI202010411}, making them highly advantageous for a wide range of applications, including impact protection \cite{D4SM01144A,Gu2023}, vibration damping \cite{Zhang2008}, and adaptive materials \cite{Wei2018}. Recent studies on sandwich structures have demonstrated that embedding STFs in composite cores can markedly enhance vibration isolation and even add functionalities such as electrical conductivity \cite{Sheikhi2023Isolation,Sheikhi2023MSTF}. Unlike actively controlled systems such as magnetorheological (MR) \cite{Jolly1999, Ginder1996} and electrorheological (ER) dampers \cite{Stanway1996, Wereley2008}, STF-based systems function passively and need no external power \cite{Gu2023, Gurgen2018}. This inherent simplicity yields faster response times \cite{Sztandera2021} and reduced system complexity, essential for effective vibration control in environments with fluctuating ambient conditions.

In contrast to Newtonian fluids, where viscosity remains constant, leading to a linear relationship between shear stress and velocity gradient, STFs exhibit a nonlinear response, with viscosity increasing as the shear rate rises \cite{Barnes1989}. Consequently, at high velocity gradients, STFs experience a significant increase in shear stress and reaction forces, whereas these forces diminish when the velocity gradient approaches zero. A critical question arises when STFs are subjected to harmonic excitation, where the velocity gradient oscillates between low and high values. Understanding the response of STFs to harmonic forcing is essential for the effective design of vibration control devices that leverage STF properties.
Recent optimisation frameworks that link STF characteristics to rheological targets now enable researchers to tailor STF formulations for specific dynamic applications \cite{Sofuoglu2024}. The shear-thickening behavior of STFs has been attributed to several key theoretical mechanisms \cite{WEI2022110570}:

\begin{itemize}
    \item \textbf{Order-disorder transition (ODT):} At low shear rates, particles form ordered structures. As the shear rate increases, these structures are disrupted, leading to a sudden rise in viscosity \cite{Hoffman1975, Bender1996}.
    \item \textbf{Hydro-clustering:} Shear stress induces the formation of transient particle clusters, increasing the effective viscosity of the fluid \cite{Wagner2009, WOS:000240958300006}.
    \item \textbf{Jamming:} At high particle concentrations, interparticle interactions hinder flow, causing the system to transition into a jammed state or quasi-solid \cite{PhysRevE, Romcke2021}.
    \item \textbf{Frictional contact:} At elevated shear rates, lubrication effects between particles decrease, leading to increased frictional contacts and normal forces, further amplifying viscosity \cite{Pan2017, jamali2019alternative}.
\end{itemize}

While existing mechanisms shed light on the transition of shear-thickening fluids (STFs) from a liquid-like to a solid-like state under shear stress, they do not fully account for the nonlinear behavior observed under normal stress. In contrast, Newtonian fluids typically exhibit linear dynamics that are easier to predict and analyze \cite{Zamanian2015discretization}. However, analyzing nonlinear dynamics poses significant challenges \cite{rezaei2017two}, and STFs naturally tend toward such complex behavior. This nonlinearity, especially under harmonic normal stress excitation, often results in irregular fluctuations \cite{REZAEI2023104503}.

In many practical applications, such as adaptive damping systems, STFs experience cyclic normal stress conditions. Whilst numerous works have explored STF responses to shear stress and impact loading, their behavior under harmonic normal force excitation remains poorly understood. Lin et al. \cite{Lin_2019} investigated STF dampers under long-term cyclic loading at various constant excitation frequencies, revealing that performance depends on excitation frequency. However, their study did not clearly define the relationship between the applied frequencies and the system’s resonance frequency, a critical factor influencing damper performance. Moreover, in practical applications, excitation frequency often varies over time, and systems rarely undergo prolonged loading cycles.

To our knowledge, all previous experimental studies focused on shear-mode loading at fixed frequencies and did not examine how STF behavior evolves during a continuous frequency sweep. Our study is the first to apply harmonic normal loading with a frequency sweep, which uncovers three previously unreported STF regimes.

Here, we address these gaps by examining STF behavior under harmonic normal stress excitation with continuously varying frequency. We analyze the STF response across a continuous frequency spectrum—before, during, and after resonance—providing essential insights for optimizing vibration control systems. To achieve this, we conducted a series of experiments utilizing an electric motor coupled with an unbalanced mass to generate harmonic excitation forces. These forces were transmitted to a rectangular plate submerged in an STF contained within a stationary container. Initially, the system was characterized without the STF to facilitate parametric identification of the baseline oscillator model \cite{Kudra2023}. Subsequent experiments involving the STF enabled us to compute the fluid’s force response and compare it with recorded experimental data.

Section~\ref{Experimental Stand} describes the experimental study. Initially, the system was analyzed without STFs to perform parametric identification, enabling the modeling of the oscillator, as explained in section~\ref{Physical and Mathematical Modeling of the Oscillator}. Subsequently, experiments were carried out with STFs, and the identified system parameters were used to compute the STF force response, which was then compared with the recorded data in section~\ref{STF Force Based on Parameter Identification}. In Section~\ref{Results and Discussion}, experimental data were examined under varying excitation frequencies, both at constant rates of increase and at fixed values. The excitation frequency, displacement response, and fluid force reaction were recorded over time. The collected data were used to present time-history responses of displacement, velocity, acceleration, STF force, hysteresis plots, and the frequency response of the system. Through comparative analysis, the unique dynamic behavior of STFs under harmonic excitation was identified and discussed. Finally, section~\ref{Conclusion} presents the key findings and their implications.

This study contributes to the growing body of knowledge on STF dynamics and supports the advancement of smart, adaptive damping systems with tunable mechanical properties and energy-efficient performance.

\section{Experimental stand} \label{Experimental Stand}
\subsection{Materials}\label{Materials}
The shear-thickening fluid (STF) used in the experiments was carefully prepared using the following components. It is worth mentioning that, since this study focuses on the dynamic response of STF rather than the rheological properties of STF suspensions, the starch was used as received, without oven-drying \cite{Bokman_2022}. The required amount of water was measured, and the chemical agent (Triton (R) X-100) was added to create a uniform solution. Gradually, the starch was added while continuously stirring to avoid clumping, ensuring a homogeneous suspension; the fully mixed STF was then left to stand for approximately two hours at room temperature. The exact quantities of the components are listed in  Table~\ref{tab:STFComponents}.
\begin{table}[ht]
    \centering
    \caption{Components used to prepare the STF}
    \label{tab:STFComponents}

    \newcommand{\hdr}[2]{\shortstack[c]{\rule{0pt}{2.6ex}#1\\[0.4ex]#2}}

    \begin{tabular}{|l|c|c|c|c|}
        \hline
        \textbf{\hdr{Component}{}} &
        \textbf{\hdr{Quantity}{(g)}} &
        \textbf{{\hdr{wt}{(\%)}}} &
        \textbf{{\hdr{Density}{(g/cm$^{3}$)}}} &
        \textbf{{\hdr{$\phi_{\mathrm{eff}}$}{(vol\%)}}} \\ \hline
        \rule{0pt}{2.3ex}Potato starch &
        381.49 & {63.35} & {1.50} & {53.66} \\ \hline
        \rule{0pt}{5ex}\shortstack[c]{Chemical agent\\(Triton\textsuperscript{\textregistered}\,X‑100)} &
        \hdr{15.83}{} & {\hdr{2.63}{}} & {\hdr{1.07}{}} & {\hdr{3.12}{}} \\ \hline
        \rule{0pt}{2.3ex}Water &
        204.88 & {34.02} & {1.00} & {43.22} \\ \hline
    \end{tabular}
\end{table}

The term \textbf{wt\%} indicates each component's mass percentage in the total mixture, while \textbf{$\phi_{eff}$ (vol\%)} is the volume fraction each component occupies in the suspension. Previous studies have conducted detailed rheological analyses of STFs, including formulations closely matching the one used in our research. Maharjan and Brown \cite{Maharjan2017} investigated a cornstarch-in-water suspension nearly identical in composition to ours, and includes viscosity–shear rate curves, critical shear-thickening onset, normal stress evolution, and yield stress in the jammed state. Waitukaitis and Jaeger \cite{Waitukaitis2012} examined a dense cornstarch suspension that is also comparable to ours, reporting steady-state rheology and dynamic jamming front behavior under impact. Egawa and Katsuragi \cite{Egawa2019} studied several dense potato-starch suspensions, one of which closely resembles our formulation, and provided rebound-derived elastic modulus and the viscosity ratios. However, these studies did not explore the system response of such STFs under harmonic excitation forces. The present study addresses this gap by investigating the behavior of a potato starch-based STF under harmonic excitation, providing new insights into its dynamic properties. The mixture was then left to rest for several hours to allow the suspension to stabilize, ensuring the shear-thickening behavior during the experiments.

\subsection{Description of the Experimental Setup} \label{Description of the Experimental Setup}
\begin{figure}
    \centering
    \includegraphics[width=0.8\textwidth]{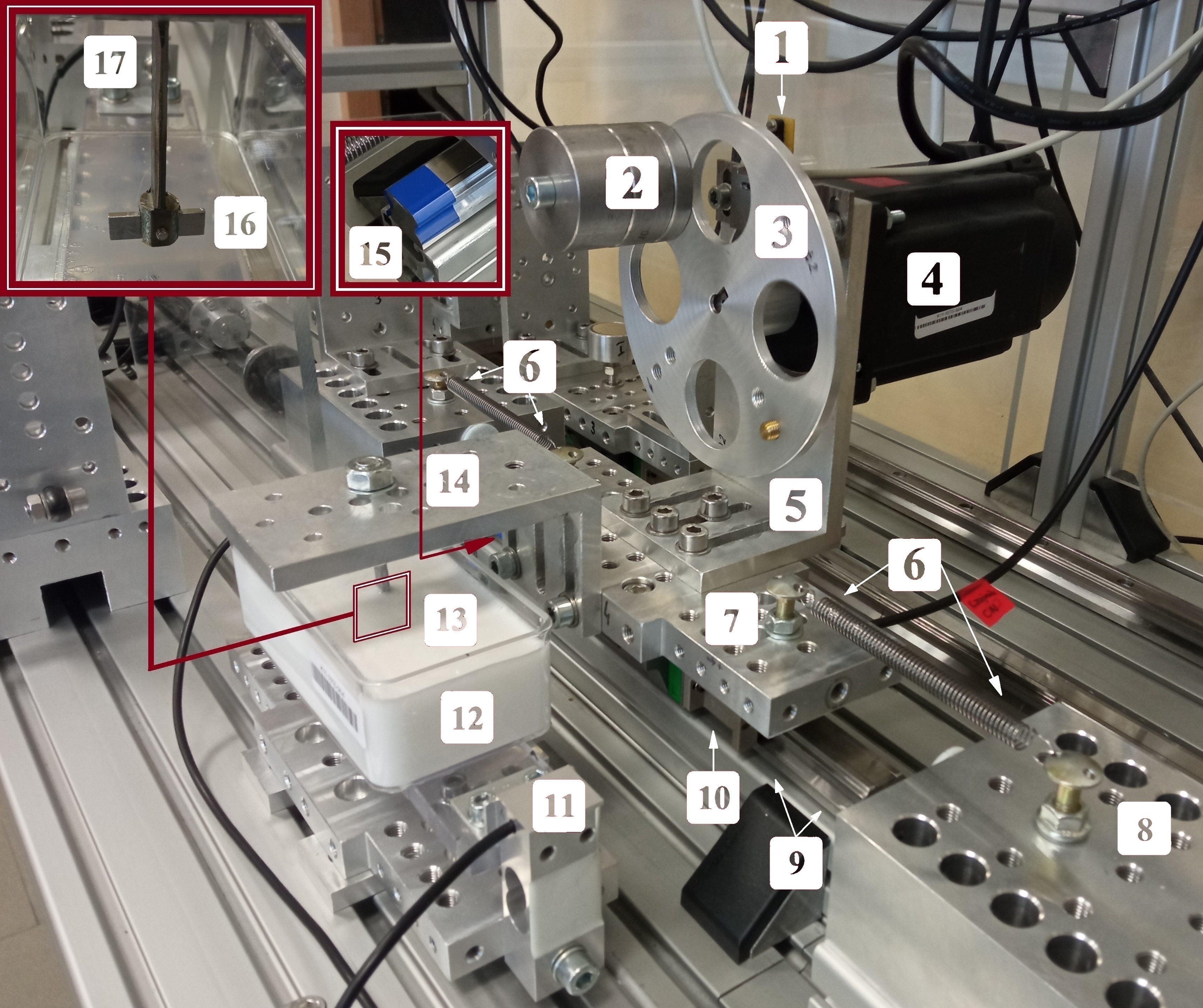} 
   \caption{Experimental setup with labeled components.}
    \label{fig:ExperimentalSetup}
\end{figure}
The experimental setup, illustrated in Fig.~\ref{fig:ExperimentalSetup}, is based on a movable, inertia-driven oscillator with a single degree of freedom. The main components of the setup are:

\begin{enumerate}
    \item \textbf{Optical interrupter (1)}: Measures the angular position ($\phi$) of the unbalanced mass.
   \item \textbf{Unbalanced mass (2)}: With a mass of \( m_0 = 0.77835 \) kg, it is mounted at radius \(e = 0.6\,\text{m}\), generating harmonic excitation forces.
    \item \textbf{Disk (3)}: Driven by the motor, it holds the unbalanced mass.
     \item \textbf{Motor (4)}: Powered by a rotary encoder, controlled through a closed-loop system.
    \item \textbf{Motor mounting bracket (5)}: Secures the motor to the trolley.
    \item \textbf{Mechanical springs (6)}: Stabilize the trolley by connecting it to the frame.
    \item \textbf{Trolley (7)}: The primary moving part of the system.
    \item \textbf{Frame (8)}: Provides stationary structural support for the entire system.
    \item \textbf{Profile rail (9)}: Guides the movement of the trolley.
    \item \textbf{Hall displacement sensor (10)}: Measures the trolley’s displacement using the magnetic tape (15).
 \item \textbf{Strain-gauge sensors (11):} Measure the resistance forces produced by STF using two BTENS-L602JM-060-C2 strain gauges.

\begin{enumerate}
\sloppy
    \item \textit{ Sensor Placement:} The sensors were bolted between the container’s stainless-steel base and upper support blocks, aligned with the force direction, and connected in 4-wire differential mode to an NI 9361 card.
    
    \item \textit{Calibration Procedure:} Calibration was performed before each test using an AXIS FB-500 dynamometer as the reference sensor. Two BTENS-L602JM-060-C2 full-bridge strain gauges were bolted between the stainless-steel base and the support blocks. A preload was applied to the elastic element to eliminate initial offset. The force measured by the dynamometer was compared with the voltage output of the strain gauge system. The load was increased in increments of 1 kg, and the strain gauge constants were updated after each step. The output signal was recorded in LabVIEW and fitted using a linear function: \( y = a \cdot x + b \), where \( a = 0.2942906 \) and \( b = 0.0044168 \). These calibration parameters were approximately validated by comparing the tension applied to the dynamometer with the processed LabVIEW output.
\end{enumerate}
    \item \textbf{Container (12)}: Stationary container that holds STF.
   \item \textbf{STF (13)}: Fills the container and interacts with the mixing blade, significantly influencing the system response.
 \item \textbf{Clamp (14)}: Transfers the force reaction from the mixing blade to the trolley.
     \item \textbf{Magnetic tape (15)}: Attached to the front of the profile rail for displacement measurement.
         \item \textbf{Mixing blade (16)}: Fully submerged in the STF, ensuring a minimum distance from the container walls equal to the mixing blade’s width. It interacts with the STF with an immersion depth of 35 mm from the top edge of the container, with a cross-sectional area of approximately \( 285 \, \text{mm}^2 \). 
   \item \textbf{Mixer connector (17)}: Links the clamp (attached to the trolley) to the mixing blade, with (rounded) beveled edges to reduce movement resistance.
\end{enumerate}

This experimental configuration allows the mixing blade (16) to move reciprocally within a stationary container (12) filled with STF (13).
The stationary container (12) has the following geometric dimensions:
\begin{itemize}
    \item Length: 145 mm,
    \item Width: 70 mm,
    \item Height: 45 mm.
\end{itemize}
The trolley (7) moves along the profile rail (9), with a magnetic tape (15) on its opposite side, which interacts with the Hall sensor (10) to track displacement. The trolley (7) is stabilized by mechanical springs (6), which connect it to the frame (8). The motor (4), mounted on a bracket (5) attached to the trolley (7), is equipped with a rotary encoder and operates through a closed-loop system. LabVIEW controls the motor. It sends the desired waveform through an NI 9361 card to a SIGLEND SDG-1025 square-wave generator.

The motor drives the disk (3), which holds an unbalanced mass (2), thereby generating harmonic excitation forces. The angular position of the unbalanced mass (\(\phi\)) is detected by an optical interrupter (1). The resulting unbalanced force, \( m_o \cdot e \cdot \omega^2\cdot \sin(\phi t) \), is transferred from the unbalanced mass (2) through the clamp (14) and the mixer connector (17) to the mixing blade (16), which interacts with the STF (13) in the container.

Strain gauge sensors (11) are placed on both sides of the container (12) to measure the generated resistance forces.

Real-time data collection in the experiment captures the following three types of data:
\begin{enumerate}
    \item The angular position over time (\(\phi\)), along with the corresponding angular frequency or rotor speed ($\omega$), is measured by the rotary encoder integrated with the motor, representing the excitation frequency. 
    \item System displacement over time $x$, measured by the Hall sensor at position (10).
    \item Resistance force of the STF over time $F_{STF}$, recorded as voltage signals from the strain gauge at position (11).
\end{enumerate}

\subsection{Data filtration} \label{Data_Filtration}

All collected data are processed and exported using LabVIEW software. To ensure accurate and reliable data for subsequent analyses, a data filtration process was implemented using MATLAB. This process addressed noise and artifacts in the raw measurements and was tailored for rotation speed, displacement, and force data.

\begin{enumerate}  
    \item \textbf{Rotation speed data filtering:}  
    \begin{itemize}  
        \item \textbf{Differentiation of angular position:} The rotation speed (\(n_p\)) was computed by differentiating the angular position data (\(\phi\)) using MATLAB's polynomial fitting function (\texttt{polyfit}) with a second-order polynomial (\(N_p = 2\)).  
        \item \textbf{Windowed polynomial fitting:} A fitting window of \(N_d = 800\) data points was applied in MATLAB to smooth the differentiation process while minimizing noise amplification.  
        \item \textbf{Boundary handling:} For data points near the boundaries of the dataset, where the fitting window was incomplete, the rotation speed was set to zero to avoid inaccuracies.  
    \end{itemize}  

    \item \textbf{Displacement data filtering:}  
    \begin{itemize}  
        \item \textbf{Velocity and acceleration computation:} Displacement data (\(x\)) was processed in MATLAB to compute velocity (\(v\)) and acceleration (\(a\)) using \texttt{polyfit} with a third-order polynomial (\(N_p = 3\)) over a smaller window of \(N_d = 40\) data points.  
        \item \textbf{Noise reduction:} The polynomial fitting method in MATLAB effectively reduced high-frequency noise while providing smooth derivatives, ensuring accurate calculations of system dynamics.  
        \item \textbf{Boundary values:} For dataset boundaries, where polynomial fitting was not feasible, velocity and acceleration values were set to zero to maintain consistency in the results.  
    \end{itemize}  
\end{enumerate}  

The MATLAB-based filtration process ensured high fidelity in rotation speed, velocity, and acceleration data. Filtered data were stored as MATLAB files (\texttt{.mat}) for further analysis, enabling precise modeling and experimental validation.  

\section{Physical and mathematical modeling of the oscillator} \label{Physical and Mathematical Modeling of the Oscillator}

\begin{figure}
    \centering
    \includegraphics[width=\textwidth]{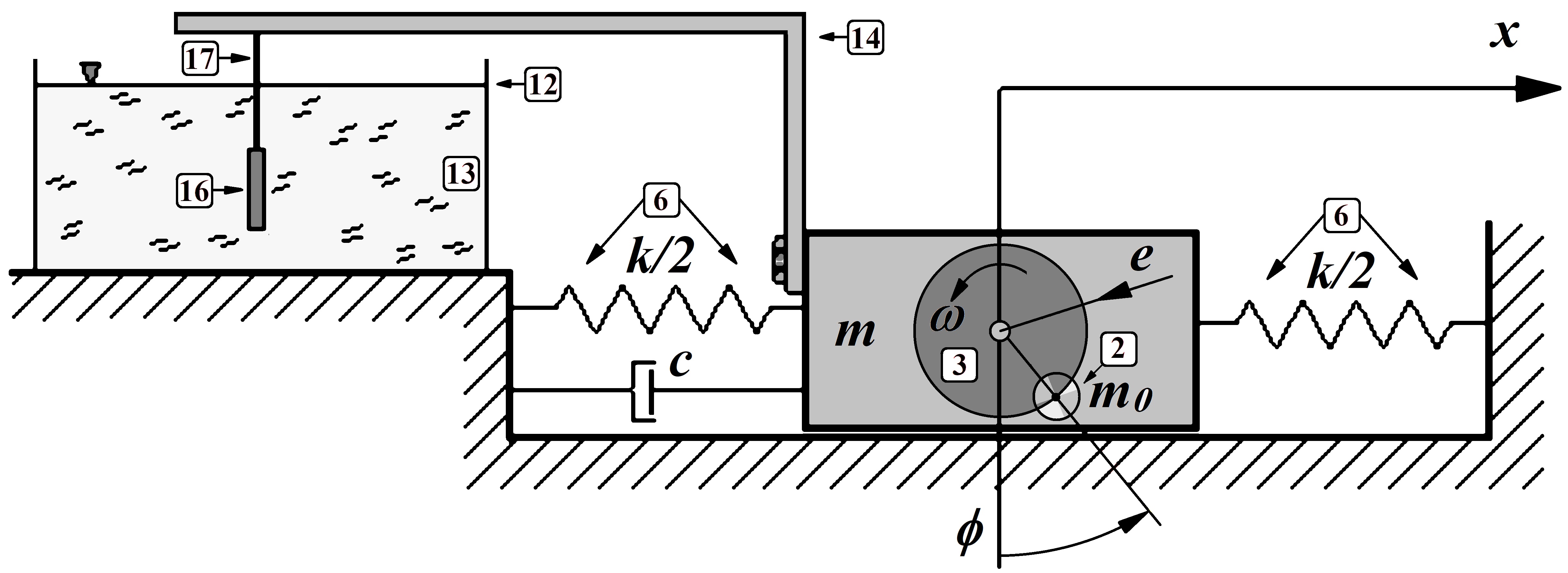} 
    \caption{Physical model of the oscillator.}
    \label{fig:physical_model}
\end{figure}

Fig.~\ref{fig:physical_model} shows a simplified physical model of the experimental setup described in Subsection~\ref{Description of the Experimental Setup}. The component labels in this figure are the same as those used in Fig.~\ref{fig:ExperimentalSetup}. The system is modeled as a movable, inertia-driven oscillator with one degree of freedom and a total mass \( m \).
 The system’s oscillating mass, the cart, moves along the $x$-axis and is connected to the base frame via springs with a total stiffness $k$. At the equilibrium position ($x=0$), the forces in the springs balance each other. The system includes damping, modeled by a coefficient \( c \), which accounts for linear losses in the rolling bearings.

As shown in the figure, an unbalanced mass \( m_0 \) is mounted on a disc at a radius \( e \) (the eccentricity) with an angular position \( \phi \). The excitation angular velocity, or rotor speed, is defined as \( \omega = \dot{\phi} \), resulting in a harmonic excitation force ($F_0 \omega^2 \sin(\phi)$).
\subsection{Mathematical Model of the Oscillator}

Based on the physical model, the mathematical model of the oscillator shown in Fig.~\ref{fig:physical_model} is formulated as:

\begin{eqnarray}\label{eq:MathematicalModel}
    m \ddot{x} + k x + F_R(\dot{x}) + F_{STF} = F_0 \omega^2 \sin(\phi),
\end{eqnarray}
where $F_{STF}$ represents the unknown force response of the shear-thickening fluid (STF). The magnitude of the excitation force $F_0$ is given by:
\begin{equation}
    F_0 = m_0 e.
\end{equation}

Additionally, \( F_R(\dot{x}) \) is the resistance force in the linear rolling bearing.

\subsection{Resistance Force Model}

The resistance force $F_R(\dot{x})$ is modeled as:

\begin{eqnarray}\label{eq:ResistanceForce}
    F_R(\dot{x}) = c\dot{x} + \left(T + \Delta T \frac{1}{1+\eta \lvert \dot x \rvert}\right) \frac{\dot x}{\sqrt{{\dot x}^2+{\epsilon}^2}},
\end{eqnarray}

Where:
\begin{itemize}
    \item $c$ is the linear damping coefficient,
    \item $T$ is the magnitude of the dry friction force at high velocity,
    \item $\Delta T = T_0 - T$ is the difference between the static dry friction force $T_0$ and $T$,
    \item $\eta$ controls the speed of decreasing the dry friction force, from \( T_0 \) to \( T \), as velocity increases,
    \item \( \frac{\dot{x}}{\sqrt{\dot{x}^2 + \epsilon^2}} \) is a smooth approximation of the sign function, where \( \epsilon \) is a small numerical parameter.
\end{itemize}

Although the system involves linear rolling bearings rather than classical dry friction, their resistance behavior is effectively captured by the hybrid model in Eq.~(\ref{eq:ResistanceForce}). This model, which combines linear damping with a smoothed Coulomb friction term, has been successfully applied in our previous studies on the same cart-rail system in a similar range of speeds~\cite{Kudra2023,Witkowski2021MSSP, Witkowski2022MSSP,Kudra2022MSSP} and other rolling contacts~\cite{Perlikowski2022MSSP}, making it a simple and reliable choice for modeling rolling resistance in our setup.

\sloppy To ensure smoothness in the nonlinear damping term, we introduce the parameter $\epsilon$, which prevents numerical instability as the velocity $\dot{x}$ approaches zero. To avoid truncating the friction characteristic peak at its maximum ($T_0 = T + \Delta T$), we choose:

\begin{equation}
\label{eq:epsilon}
    \epsilon = 10^{-3} \eta^{-1}.
\end{equation}

This choice ensures that the friction characteristic peak is not truncated when it reaches its maximum value.

\section{Numerical simulation of the oscillator without STF} \label{Numerical simulation of the oscillator without STF}

In this section, we present the results of numerical simulations (parametric identification) and experimental investigations of the system without the STF damper, meaning \( F_{\text{STF}} = 0 \). The parameter estimation method is based on minimizing the objective function, which represents the sum of squared differences between experimental and simulated data:
\begin{equation}
\label{eq:ObjectiveFunction}
S(\mu) = \frac{1}{N}\sum_{i=1}^{N}\left(x_{\text{exp}}(i)-x_{\text{sim}}(i,\mu)\right)^2,
\end{equation}
where data points are recorded \( N \) times over the total time interval \( T \). Here, \( S(\mu) \) represents the average squared deviation between the experimentally measured position \( x_{\text{exp}}(i) \) of the cart (the system's oscillating mass) and the corresponding numerically simulated position \( x_{\text{sim}}(i,\mu) \), computed using the mathematical model described by Eq.~(\ref{eq:MathematicalModel}). The vector \( \mu \) contains the estimated parameters.

The initial conditions for numerical simulations are taken from the experiment at the first point of the series: the cart's position is directly measured, and its velocity is obtained via numerical differentiation. The phase of the external forcing in the numerical simulations, represented by the angle \( \phi \), is also taken from the experiment. 

To match the numerical simulation with the experimental data, we search for the global minimum of \( S(\mu) \). 
Since it is not possible to uniquely identify all parameters of the differential equation (Eq.~(\ref{eq:MathematicalModel})), we fix the total mass of the oscillator at:  

\begin{equation} \label{eq:the system’s oscillating mass}
m = 7.650 \text{ kg},
\end{equation}

As determined from precise direct measurements. Additionally, we assume a constant value for the parameter:

\begin{equation}
\label{eq:eta}
\eta = 10^3 \text{ s/m}.%
\end{equation}

Therefore, from Equation~(\ref{eq:epsilon}),  $\epsilon$ equals  $10^{-6} \, \text{m/s}$. The selected values of $\eta$ and $\epsilon$ (determined through trial simulations) ensure numerical stability and realistic smoothing of the discontinuous friction model. Values that are too small may cause solver instability or result in friction force cut-off near zero velocity, while values that are too large may excessively smooth the response and misrepresent the physical behavior. In particular, a high $\eta$ may slow the variation of the friction force with velocity. However, moderate variations around the selected values do not significantly affect the solution. 
We achieve parameters using the Nelder-Mead method, which is a direct search method for nonlinear optimization problems without derivatives. Specifically, we use the \texttt{fminsearch} function in MATLAB, which implements this method.
To support our study, we provide supplementary materials, including MATLAB codes for parametric identification using the Nelder-Mead method.

To identify the system parameters based on the model (Eq.~(\ref{eq:MathematicalModel})), we first use the experimental data from free vibrations of the oscillator to determine \( k \), \( c \), and \( T \) in subsection~\ref{Free Vibrations}. In the next subsection, subsection~\ref{Forced Vibrations}, the values from free vibrations are used as initial estimates for the Nelder-Mead method. This process refines these values and also determines additional parameters \( \Delta T \) and \( F_0 \) for the forced vibrations.

\subsection{Free vibrations} \label{Free Vibrations}

\begin{figure}
    \centering
    \includegraphics[width=\textwidth]{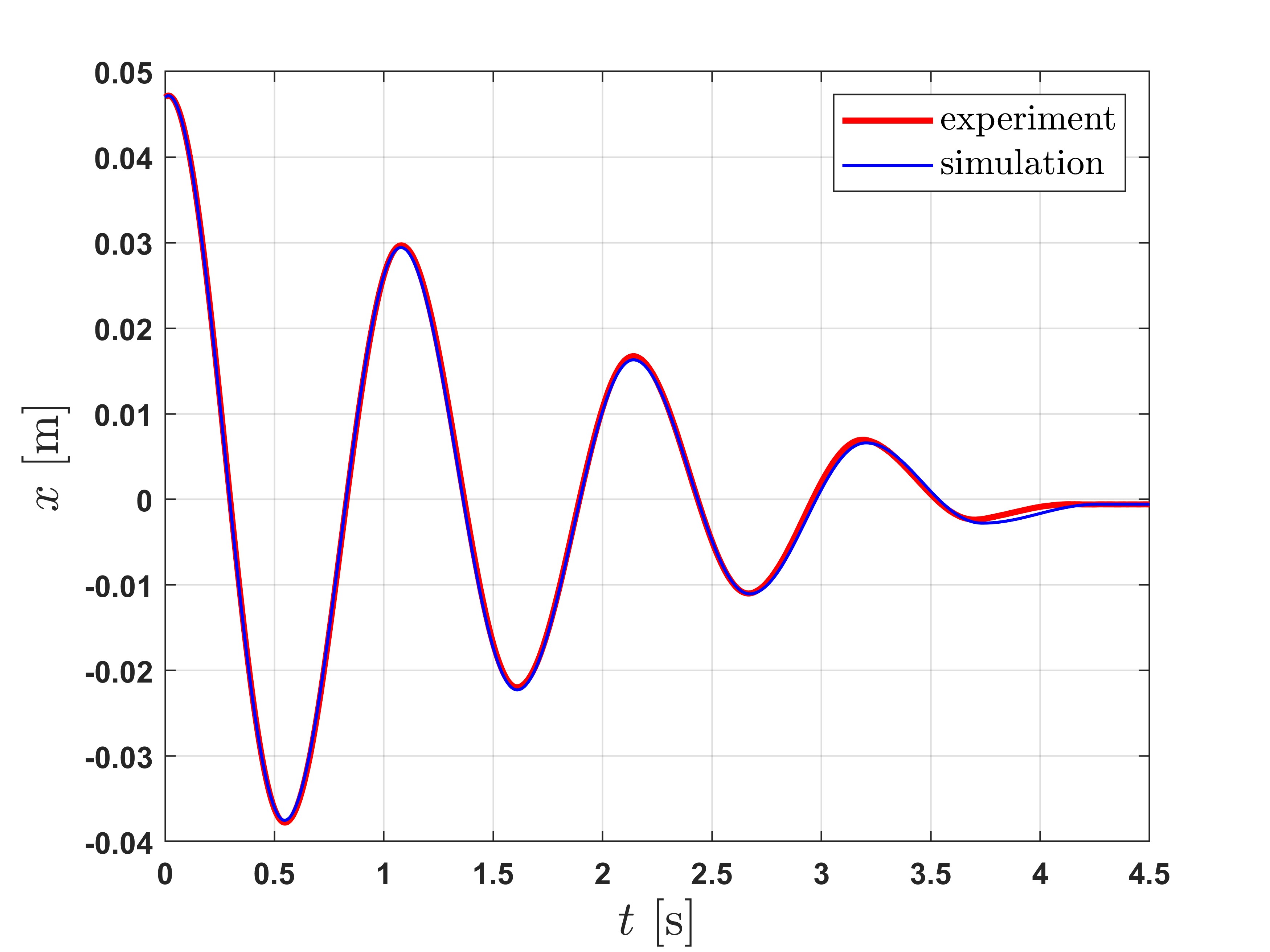}
    \caption{Comparison of experimental (red) and simulation (green) data for the free vibrations of the oscillator, showing strong agreement in both amplitude and phase.}
    \label{fig:free_vibrations}
\end{figure}

For the preliminary identification, we assume \( \Delta T = 0 \) since it has a negligible impact on the free vibration simulation. Therefore, \( \eta \) is not considered at this stage. Using this assumption, we identified the following parameters:

\begin{equation} \label{eq:FreeVibrationsPrameters}
k = 267.9245 \, \text{N/m}, \quad c = 4.3736 \, \text{N} \cdot \text{s/m}, \quad T = 0.4190 \, \text{N},   
\end{equation}

which correspond to the minimum of the objective function \( S(\mu) = 0.178 \, \text{mm}^2 \). The comparison between the numerical simulation and experimental data, using the obtained parameters, is shown in Fig.~\ref{fig:free_vibrations}.

The Fig.~\ref{fig:free_vibrations} shows a close match between the experimental and numerical results.
\subsection{Forced vibrations}  \label{Forced Vibrations}
To improve the simulation of the system's dynamics, we analyze the forced vibrations and determine the relevant parameters. This subsection presents the final parameter identification based on experimental data from the forced oscillator. The parameters obtained in subsection~\ref{Free Vibrations} (Eq.~(\ref{eq:FreeVibrationsPrameters})) are used as the starting point for the simplex method. The experiment was recorded throughout 0 to 300 seconds, with some extending to 600 seconds, with an increasing rotational speed. However, at the beginning and end of this period, the data were not smooth. In reality, the rotational speed started from zero with some fluctuations, so we excluded the initial and final portions of the data.  
\begin{figure}
    \centering
    \includegraphics[width=\textwidth]{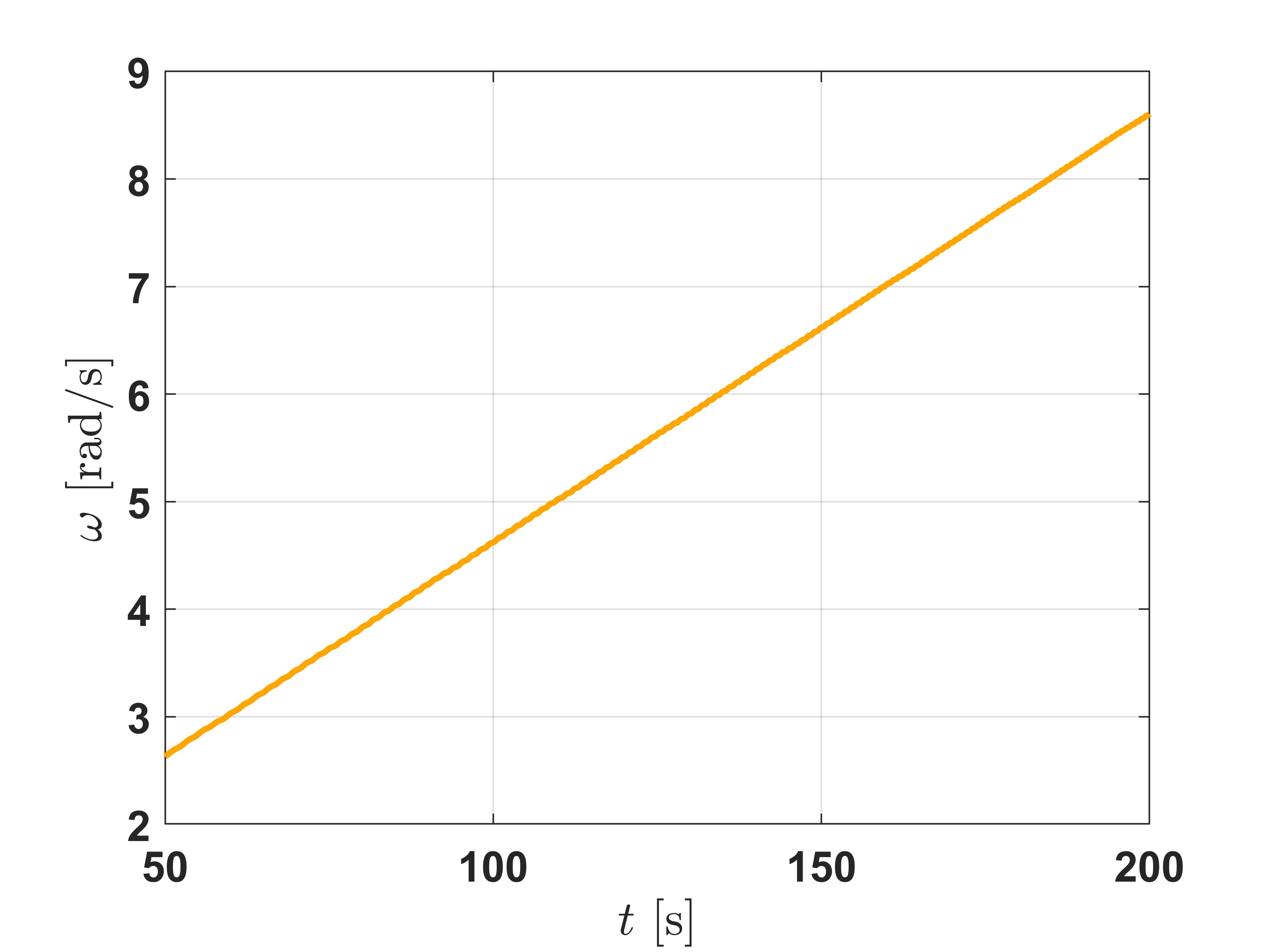} 
    \caption{Time history of the excitation frequency $\omega$ of the applied forcing, increasing at a constant rate of \( 0.047 \, \text{rad/s}^2 \) over the time interval \( t \in \langle 50, 200 \rangle \).}
    \label{fig:OmegafastIncreasing}
\end{figure}

So, based on the experimental displacement data of the oscillator at excitation frequency \(\omega\) from Fig.~\ref{fig:OmegafastIncreasing} and using the Nelder-Mead method with the same objective function (Eq.~(\ref{eq:ObjectiveFunction})), the following parameters were identified:

\begin{eqnarray} \label{eq:ForcedVibrationsPrameters}
&k = 269.9346 \, \text{N/m}, \quad c = 3.6839 \, \text{N} \cdot \text{s/m}, \quad T = 0.5424 \, \text{N}, \nonumber \\& \Delta T = 0.0947 \, \text{N}, \quad F_0 = 0.0462 \, \text{kg} \cdot \text{m}.
\end{eqnarray}

\begin{figure}
    \centering
    \includegraphics[width=\textwidth]{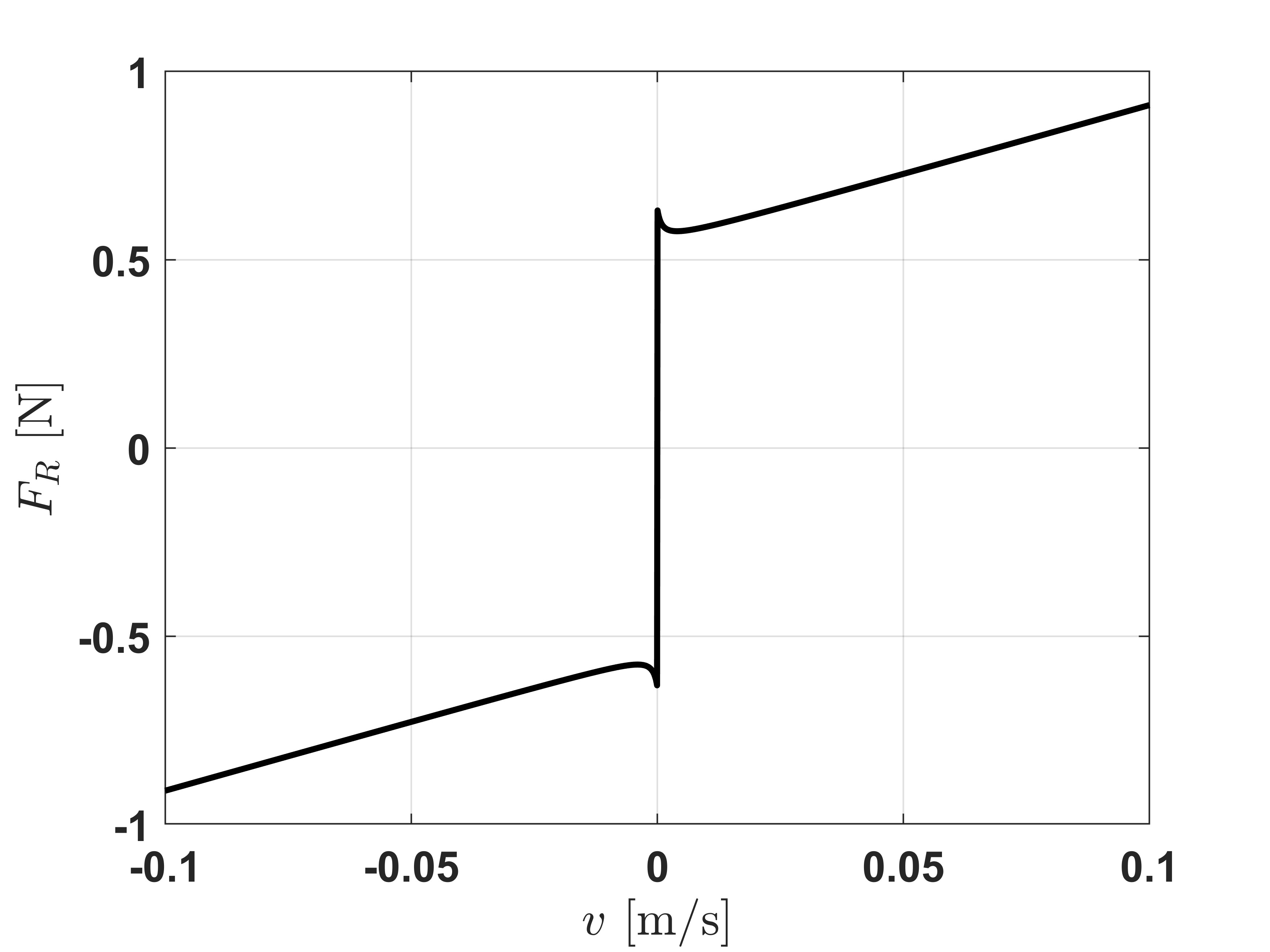} 
   \caption{Rolling bearing resistance function \( F_R (v) \) computed using the identified parameters.}
    \label{fig:ResistanceForced}
\end{figure}

These values minimize the objective function
\begin{equation}
S(\mu) = 0.567 \, \text{mm}^2,
\end{equation}
which indicates that while the error is small, it is not exactly zero. The identified parameters are valid for this system and can be applied to other conditions with varying values or increasing rates of \(\omega\) with only minor variations. With these identified parameters, the rolling bearing resistance function is plotted in Fig.~\ref{fig:ResistanceForced}.

The velocity $v$, which is the time derivative of displacement ($\dot x$), determines the rolling bearing resistance function $F_R(v)$. From this figure, it is observed that the rolling bearing resistance function $F_R(v)$ exhibits a characteristic discontinuity at $v = 0$ and demonstrates approximately linear behavior for both positive and negative velocities. The force characteristic also shows an asymmetric nature, where the resistance force increases linearly in both directions with almost the same slopes and magnitudes. This behavior resembles a modified Coulomb friction model with velocity-dependent characteristics, typical of rolling bearing systems incorporating both static and dynamic friction effects.

\begin{figure}
    \centering
    \includegraphics[width=\textwidth]{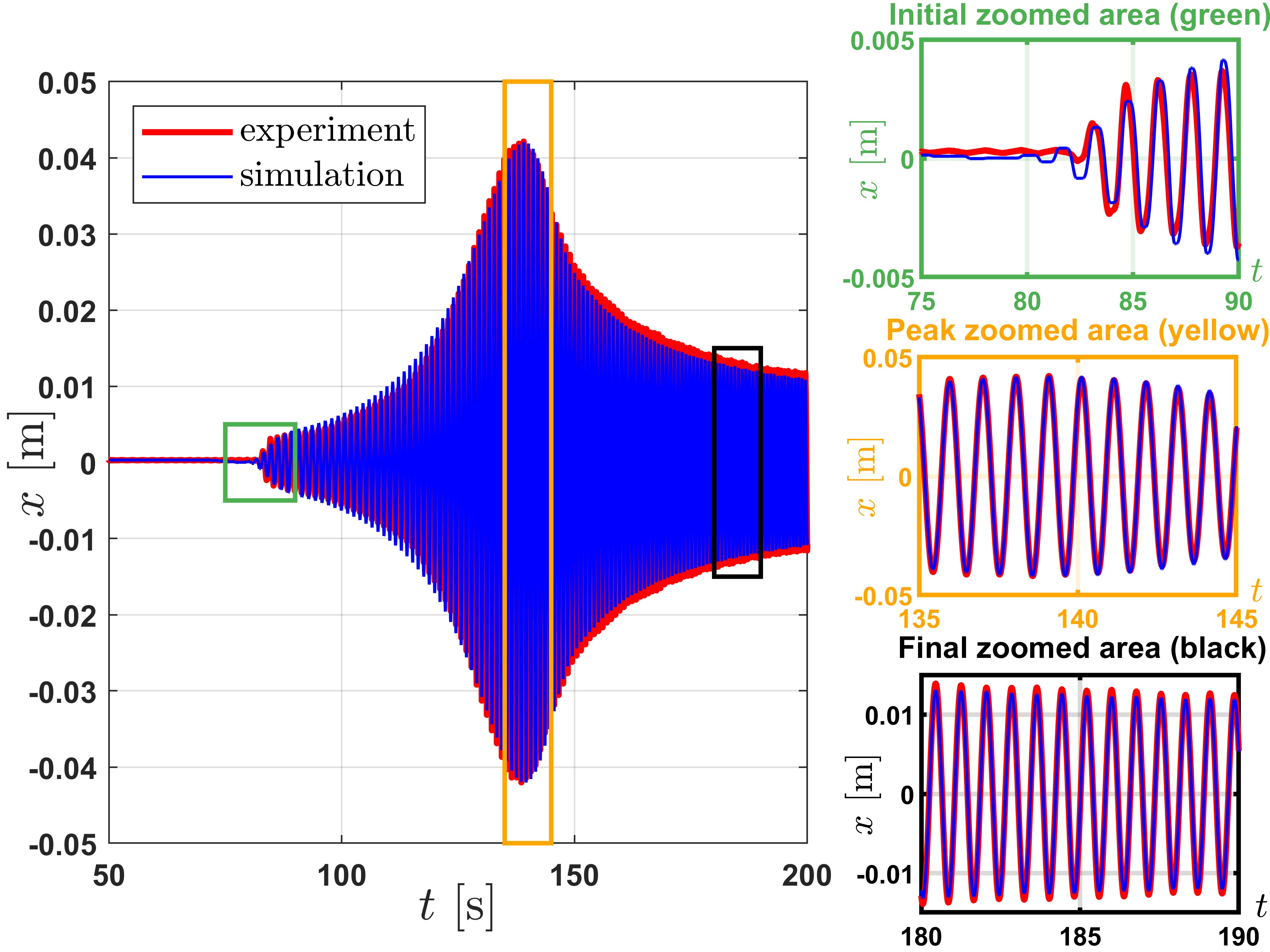} 
  \caption{Strong agreement between numerical (red) and experimental (blue) displacement ($x$) data for forced vibration over $t \in \langle 50, 200 \rangle$. Three zoomed areas are highlighted: the first (green box) from $t=75$ s to $t=90$ s, the second (yellow box) from $t=135$ s to $t=145$ s, and the third (black box) from $t=180$ s to $t=190$ s.}

    \label{fig:ForcedVibrationDisplacement}
\end{figure}

To evaluate the identified parameters, a comparison is made between the numerical simulations and experimental results. Using the parameters from Eq.~(\ref{eq:ForcedVibrationsPrameters}), displacement simulations were performed for \( t \in \langle 50, 200 \rangle \) s and compared with Hall sensor measurements ($x$), as shown in Fig.~\ref{fig:ForcedVibrationDisplacement}.
The main plot on the left demonstrates strong agreement between numerical and experimental results throughout the entire investigated time interval. To highlight the quality of this agreement at different stages of the vibration response, three critical regions are shown as magnified plots on the right: the initial zoomed area (green border) captures the onset of oscillations, the peak zoomed area (yellow border) illustrates the maximum amplitude, and the final zoomed area (black border) highlights the decay phase with decreasing amplitude. These detailed views confirm the high-quality agreement between experimental measurements and numerical simulations across all phases of the vibration response, from initial oscillations through peak amplitude to the final decay phase.

\begin{figure}
    \centering
    \includegraphics[width=\textwidth]{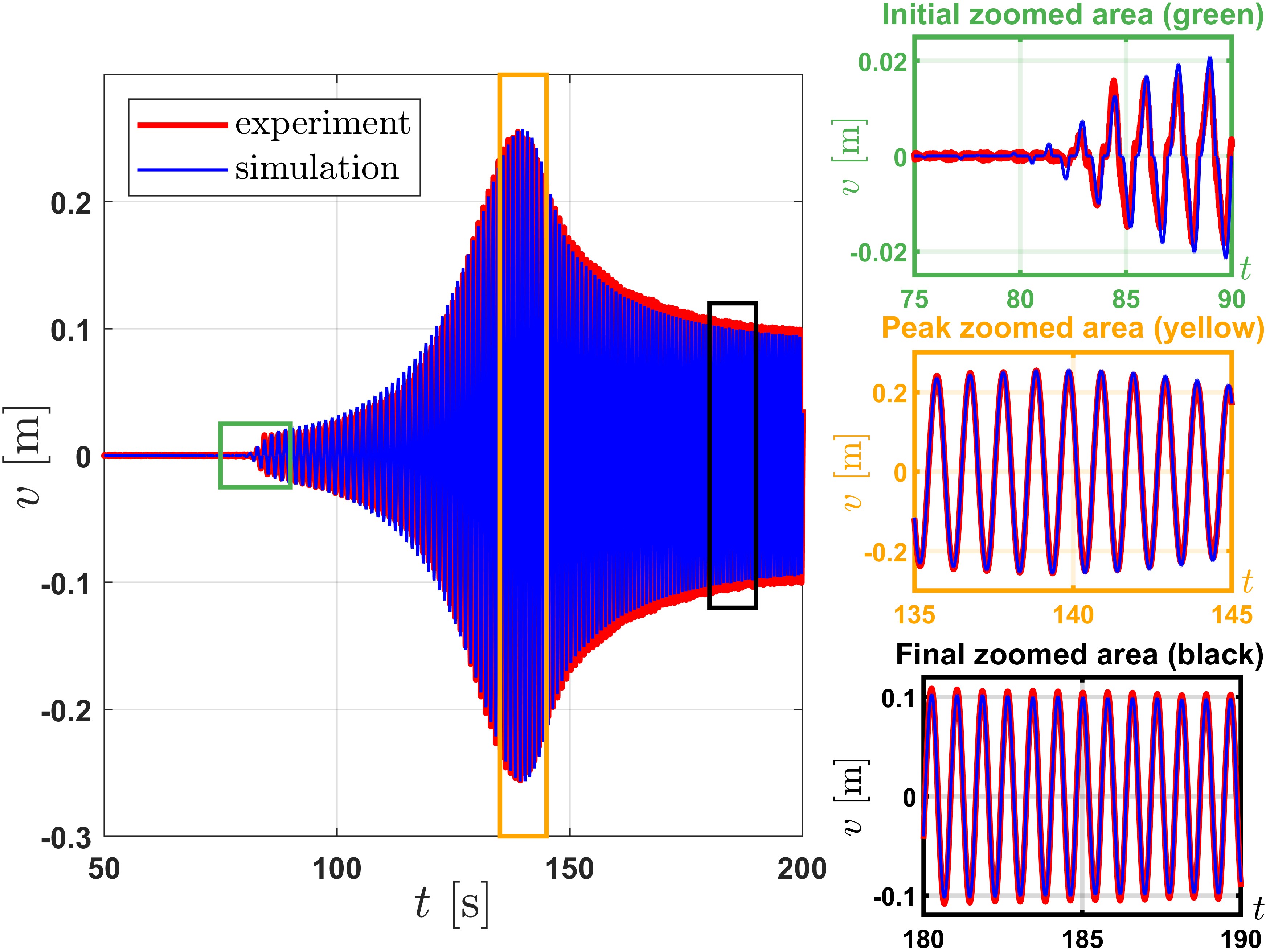} 
   \caption{Strong agreement between numerical (red) and experimental (blue) velocity ($\dot{x}$) data for forced vibration over $t \in \langle 50, 200 \rangle$. Three zoomed areas are highlighted: the first (green box) from $t=75$ s to $t=90$ s, the second (yellow box) from $t=135$ s to $t=145$ s, and the third (black box) from $t=180$ s to $t=190$ s.}

    \label{fig:ForcedVibrationVelocity}
\end{figure}

\begin{figure}
    \centering
    \includegraphics[width=\textwidth]{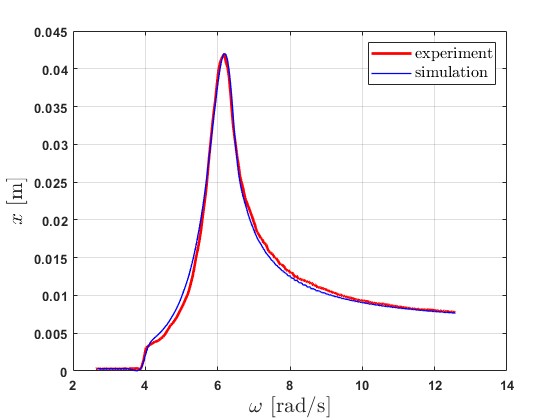} 
  \caption{Experimental (a) and numerical (b) frequency characteristics, derived from the local maxima $x_{max}$ of the displacement solutions ($x$) shown in Fig.~\ref{fig:ForcedVibrationDisplacement}.}
    \label{fig:ComparisonExperimentVsModel}
\end{figure}

An additional comparison of velocity is essential as it provides a more comprehensive verification of the mathematical simulation. Therefore, the velocity agreement between numerical simulations and experimental measurements is presented in Fig.~\ref{fig:ForcedVibrationVelocity}. The same three critical regions are magnified to demonstrate the high-quality agreement in velocity response during the initial oscillations (green border), peak response (yellow border), and decay phase (black border).
To complete the validation of our model, we present in Fig.~\ref{fig:ComparisonExperimentVsModel}, which shows strong agreement also in the frequency domain.

\section{STF force based on parametric identification} \label{STF Force Based on Parameter Identification}

In the previous section (see Section~\ref{Numerical simulation of the oscillator without STF}), experimental data without STF were used to identify the system parameters. In this section, we employ experimental data recorded with STF. As mentioned in Section~\ref{Description of the Experimental Setup}, the force of STF was also measured using a strain gauge. By rewriting Eq.~(\ref{eq:MathematicalModel}), it becomes evident that $F_{STF}$ can also be determined using the identified parameters (see Eq.~(\ref{eq:ForcedVibrationsPrameters})) and the recorded displacement $x_{exp}$ as follows: 

\begin{eqnarray}\label{eq:MathematicalModelSTFForce}
    F_{STF} = F_0 \omega^2 \sin(\phi) - m \ddot{x}_{exp} - k x_{exp} - F_R(\dot{x}_{exp}),
\end{eqnarray}

where $x_{exp}$ is the recorded displacement of the system with STF, $\dot{x}_{exp}$ is its first derivative, and $\ddot{x}_{exp}$ is its second derivative. 

The computed force due to STF can be compared with the directly recorded STF force to assess the accuracy of the identified parameters. However, numerical differentiation of $x_{exp}$ introduces visible errors and noise. Despite this, the recorded data validates the parametric identification of the system.

\begin{figure}
    \centering
    \includegraphics[width=\textwidth]{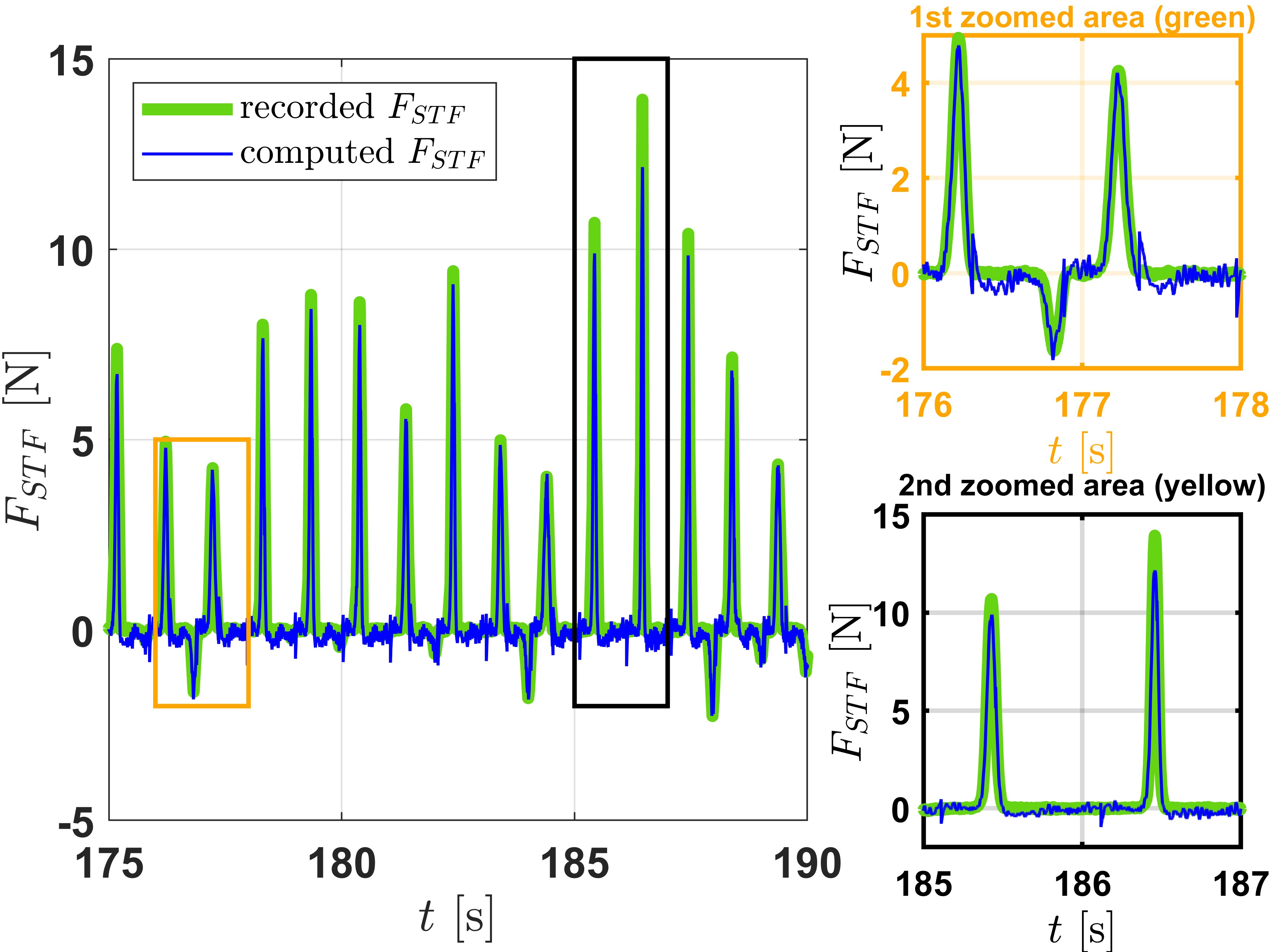}
  \caption{Comparison of the computed STF force (blue) and the recorded STF force (green)
in the critical resonance interval, $t \in \langle 175, 190 \rangle$. Two zoomed areas
are shown: the first (orange box) from $t = 176$ s to $t = 178$ s and the second (black box)
from $t = 185$ s to $t = 187$ s.}

    \label{fig:IdentifiedSTFForce}
\end{figure}

Since the parameters identified in the previous section are assumed to be valid for all excitation frequency variations, we present an additional data set recorded with a constant excitation frequency increase rate of \( 0.0333 \, \text{rad}/\text{s}^2 \). Fig.~\ref{fig:IdentifiedSTFForce} highlights a zoomed-in view of the critical interval around the resonance peak, approximately $\omega \in \langle 6, 6.5 \rangle$, as also indicated in Fig.~\ref{fig:ComparisonExperimentVsModel}. The corresponding time interval for this frequency variation is $t \in \langle 175, 190 \rangle$.

Despite the noise induced by differentiating the recorded displacement $x_{exp}$, the comparison demonstrates strong validation of our parametric identification approach.

\section{Results and discussion}
\label{Results and Discussion}

We collected data from various scenarios, including constant speeds at different levels and steadily increasing speeds at varying rates. The full dataset, along with a video, is provided in the supplementary material of this paper. To keep the presentation concise, we focus on representative samples in the following subsections.

\begin{figure}
    \centering
    \includegraphics[width=\textwidth]{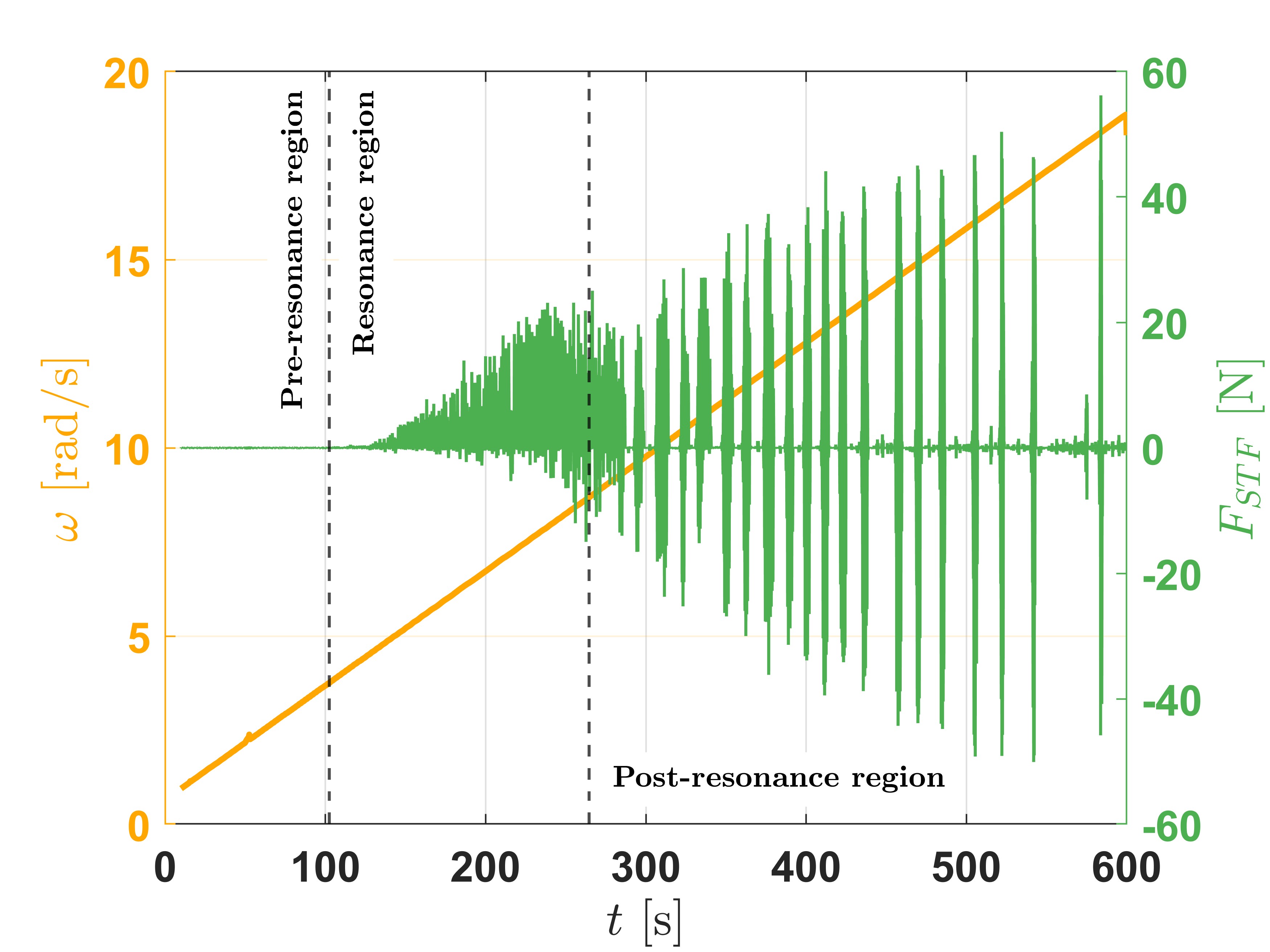}
    \caption{Time history of the excitation frequency \(\omega\) (yellow, left y-axis) increasing at a constant rate of \(0.0333\,\text{rad}/\text{s}^2\), and the STF force (green, right y-axis) over the interval \(t \in \langle 0, 600 \rangle\) seconds.}
    \label{fig:OmegaModerateIncreasing}
\end{figure}

First, to better understand \(F_{STF}\), we show in Fig.~\ref{fig:OmegaModerateIncreasing} the time histories of both the recorded \(F_{STF}\) (green, right y-axis) and the excitation frequency \(\omega\) (yellow, left y-axis), which increases at a constant rate of \(0.0333\,\text{rad}/\text{s}^2\).

From Fig.~\ref{fig:OmegaModerateIncreasing}, it is clear that \(F_{STF}\) behaves differently across three main regions:
\begin{itemize}
    \item \textbf{Pre-resonance region} (\(\omega = 0\text{--}3.75\,\text{rad/s}\)): \(F_{STF}\) remains close to zero.
    \item \textbf{Resonance region} (\(\omega = 3.75\text{--}9.22\,\text{rad/s}\)): The STF begins to respond, causing a significant increase in \(F_{STF}\).
    \item \textbf{Post-resonance region} (\(\omega > 9.22\,\text{rad/s}\)): \(F_{STF}\) exhibits impact-like behavior with growing amplitude, while the time interval between successive impacts increases.
\end{itemize}

These observations motivate a closer examination of \(F_{STF}\) in each of the three regions. Therefore, in the following subsections, we present additional figures (time histories and hysteresis plots) to further investigate the nature of the STF force. We also compare the frequency response of the system with and without STF in subsection~\ref{STF Impact: Frequency Domain Comparison with and without STF}, highlighting the influence of STF on the overall system behavior.

\subsection{Nature of STF force: time histories and hysteresis comparisons}
This subsection aims to reveal the relationship between the system's response in terms of displacement, velocity, and acceleration, and the STF force. Time-history plots of the recorded STF force and corresponding displacement are presented, along with hysteresis plots for \(F_{STF}\) versus displacement (\(x\)), velocity (\(v\)), and acceleration (\(a\)). Detailed plots for each region follow in the subsequent subsubsections.

\subsubsection{Pre-resonance region}
\begin{figure}
    \centering
    \includegraphics[width=\textwidth]{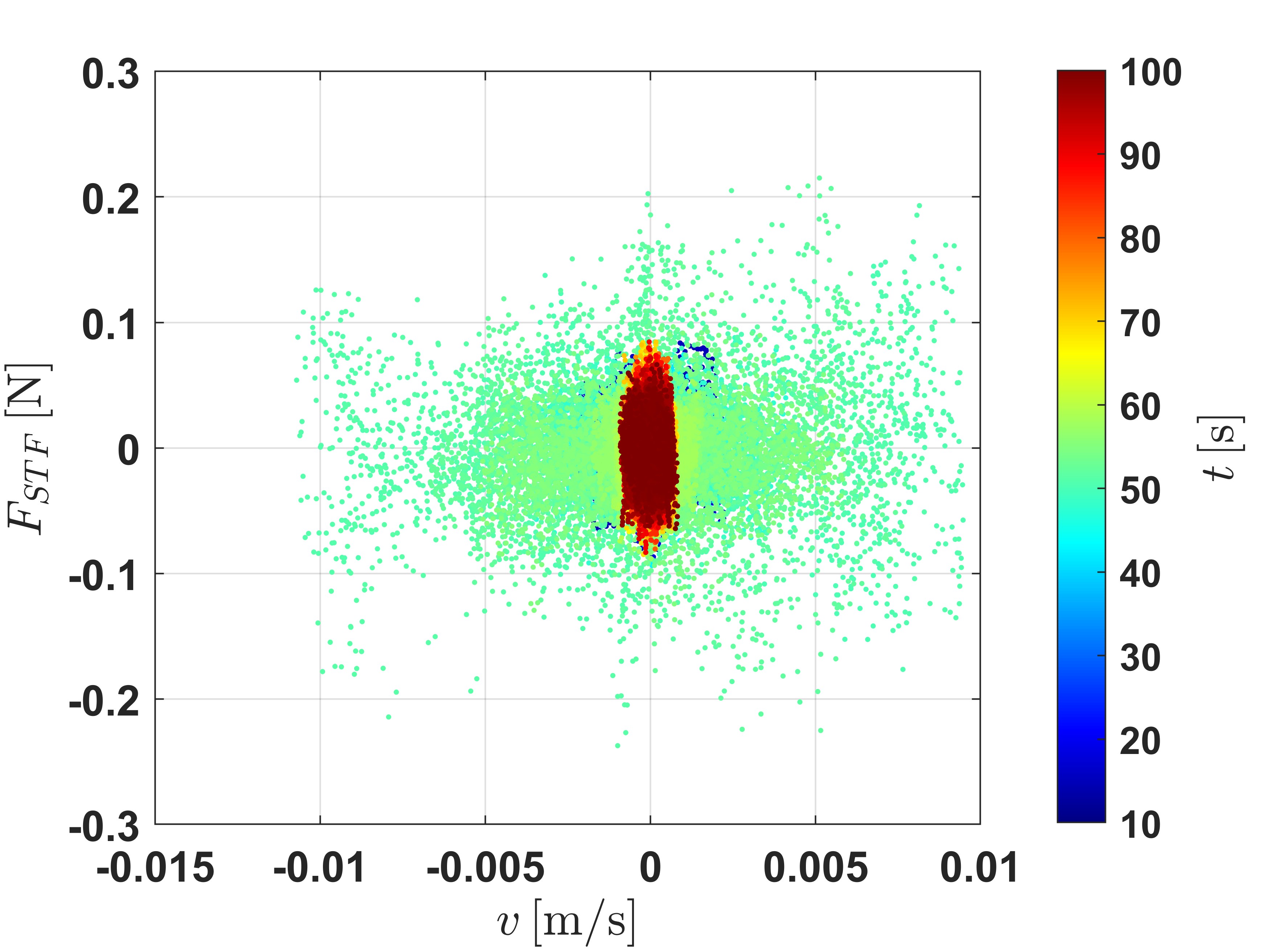} 
      \caption{Scatter plot of STF force versus velocity showing hysteresis for $t \in [10, 100]$ seconds in the resonance region. Points are colored by time, from blue ($t = 10$ s) to red ($t = 100$ s), as indicated by the colorbar.}
    \label{fig:Pre-resonance-hysteresis}
\end{figure}

As observed in the previous subsection, STF has no effect in this region, resulting in a negligible STF force ($F_{STF} \leq 0.2$ [N]). Since presenting all the plots is unnecessary, only the hysteresis plot of STF force versus displacement is shown, which displays points with almost negligible amplitude, as seen in Fig.~\ref{fig:Pre-resonance-hysteresis}.

\subsubsection{Resonance region}

To reveal the behavior of the STF force (\(F_{STF}\)) in the resonance region, we examine its relationship with the system's kinematic parameters: displacement (\(x\)), velocity (\(v\)), and acceleration (\(a\)). Specifically, we analyze how the extrema of  \(F_{STF}\) align with those of \(x\), \(v\), and \(a\). The following figures illustrate these relationships in detail.

\begin{figure}
    \centering
    \includegraphics[width=\textwidth]{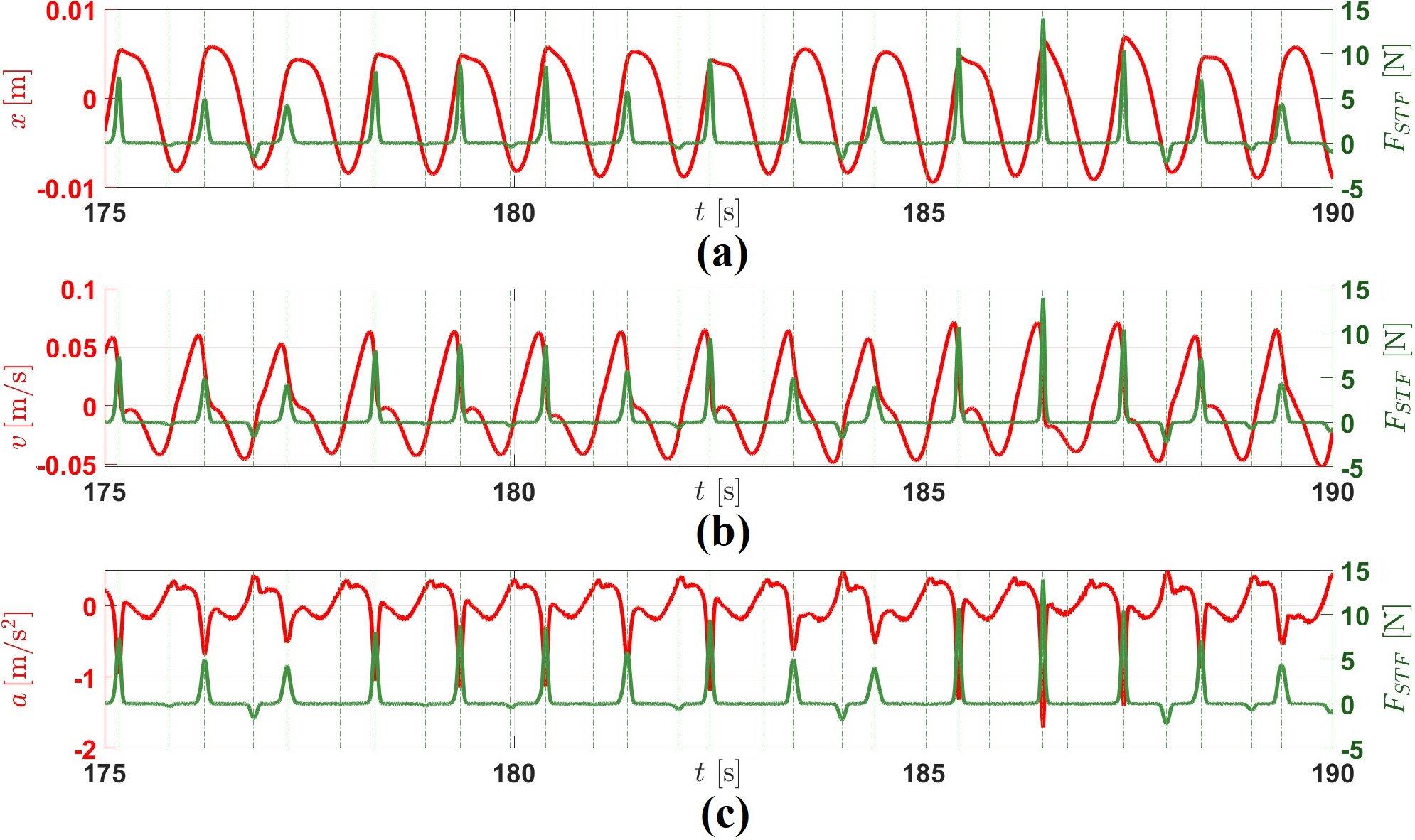}
    \caption{Time histories of (a) displacement \(x\) [m], (b) velocity \(v\) [m/s], and (c) acceleration \(a\) [m/s\(^2\)] (red, left y-axis) compared with the STF force \(F_{STF}\) [N] (green, right y-axis) during the time interval \( t \in [175, 190] \) seconds in the resonance region. The excitation frequency \(\omega\) increases at a constant rate of \( 0.0333 \, \text{rad}/\text{s}^2 \). Vertical dashed green lines indicate the extrema of \(F_{STF}\).}
    \label{fig:Resonance-TimeHistory}
\end{figure}

 From Fig.~\ref{fig:Resonance-TimeHistory}, the velocity (\(v\)) shows its peaks occurring near the zero crossings of \(F_{STF}\) with slight phase delays. Notably, the extrema of \(F_{STF}\), marked by vertical dashed green lines, align closely with those of \(x\) and \(a\), though slight phase delays are observed with \(x\). The alignment is most precise with acceleration (\(a\)), where the peaks of \(F_{STF}\) coincide exactly with the peaks and troughs of \(a\). This strong correlation suggests that \(F_{STF}\) is directly proportional to acceleration, consistent with Newton’s second law, while its relationships with \(x\) and \(v\) exhibit minor delays due to the
 system response.

\begin{figure}
    \centering
    \includegraphics[width=\linewidth]{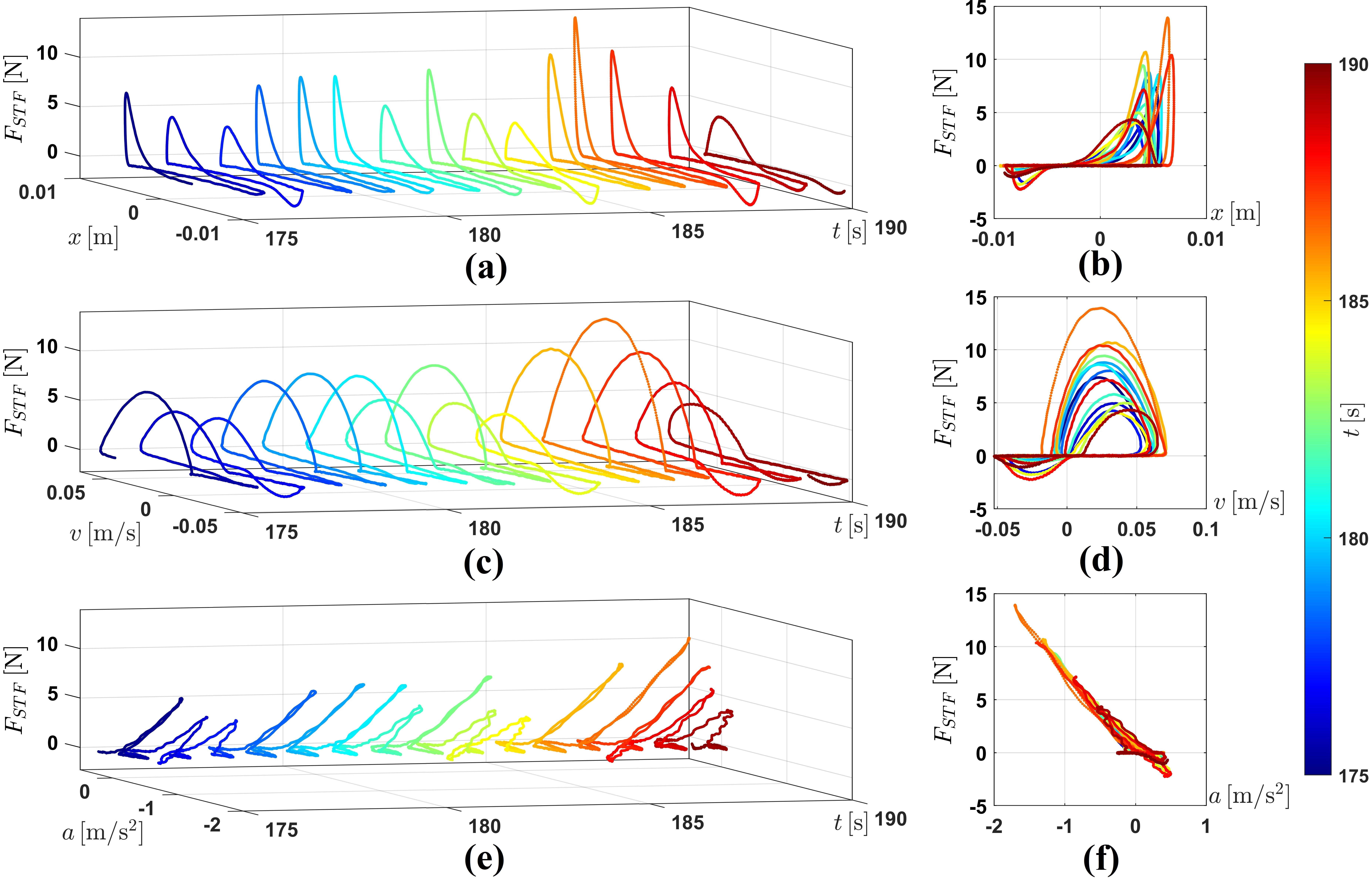}
    \caption{Hysteresis plots of \(F_{STF}\) versus (a) (3D plot) displacement \(x\) [m], (b) (2D plot) displacement \(x\) [m], (c) (3D plot) velocity \(v\) [m/s], (d) (2D plot) velocity \(v\) [m/s], (e) (3D plot) acceleration \(a\) [m/s\(^2\)], and (f) (2D plot) acceleration \(a\) [m/s\(^2\)] during the time interval \( t \in [175, 190] \) seconds in the resonance region. The lines are colored by time, from blue (\( t = 175 \) s) to red (\( t = 190 \) s), as indicated by the colorbar.}
    \label{fig:Resonance-hystersis}
\end{figure}

The hysteresis plots in Fig.~\ref{fig:Resonance-hystersis} reveal the nonlinear dynamics of \(F_{STF}\) in the resonance region. Subfigures (a) and (b) show the hysteresis loops of \(F_{STF}\) versus displacement (\(x\)) over time, with the loop areas showing irregular fluctuations over time. Such irregular behavior in the response of STFs is expected, as demonstrated in studies like \cite{REZAEI2023104503}. Subfigures (c) and (d) depict the loops for velocity (\(v\)), where the enclosed areas reflect significant energy dissipation through resonance. This increasing dissipation suggests an irregular tendency in the velocity response.  Collectively, the loop areas over the whole time within the \(F_{STF}\)-\(v\) loops highlight high-energy dissipation characteristic of this resonance region. Subfigures (e) and (f) illustrate the loops for acceleration (\(a\)), which are nearly linear with minimal enclosed areas, confirming a direct, proportional relationship between \(F_{STF}\) and \(a\).

\begin{figure}
    \centering
    \includegraphics[width=\textwidth]{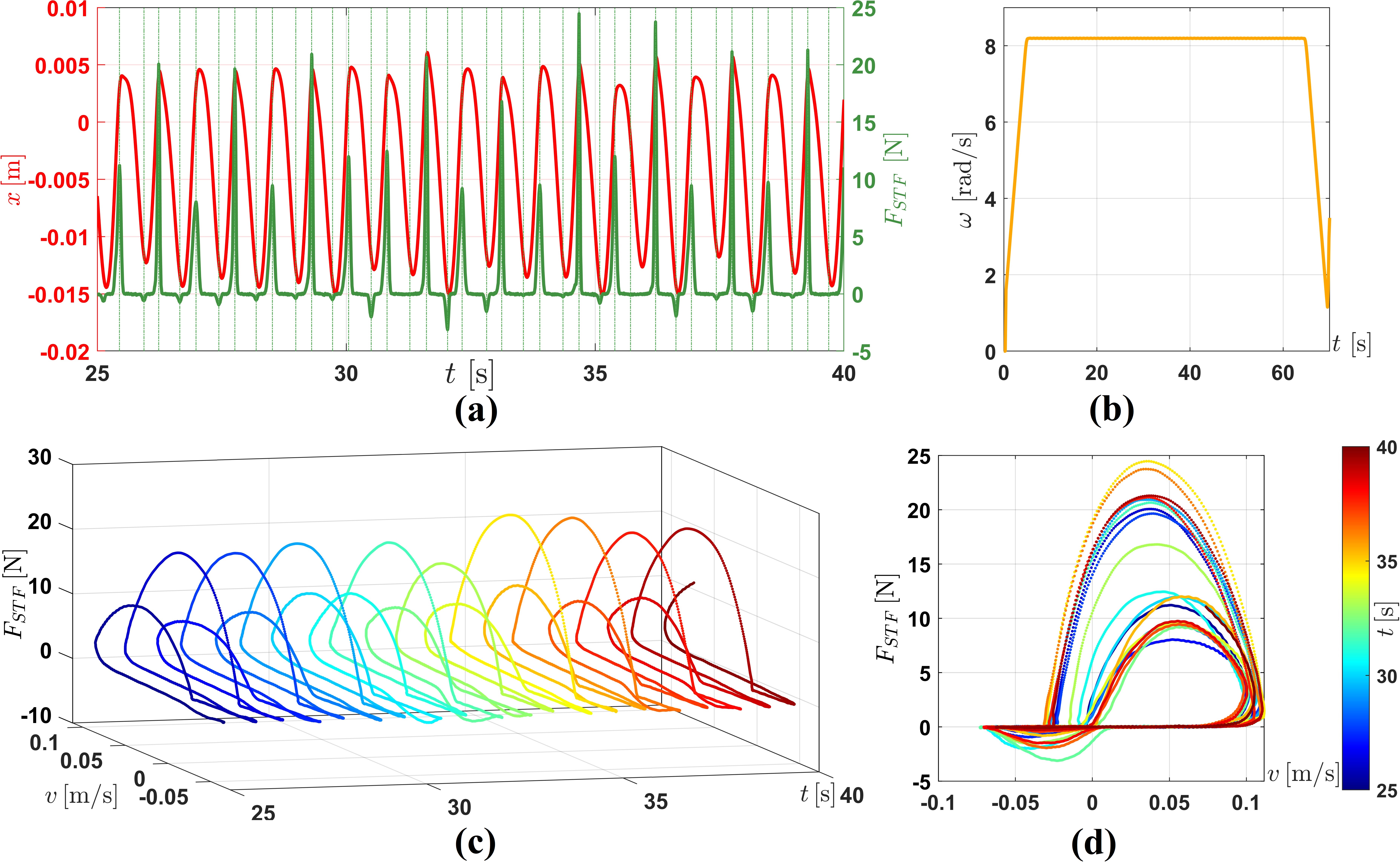}
    \caption{Plots at a constant excitation frequency \(\omega = 8.2\) rad/s: (a) Time histories of displacement \(x\) [m] (red, left y-axis) and STF force \(F_{STF}\) [N] (green, right y-axis), (b) Time history of the excitation frequency \(\omega\) [rad/s] over \( t \in [0, 70] \) seconds, (c) 3D hysteresis plot of \(F_{STF}\) versus velocity \(v\) [m/s] over time, and (d) 2D hysteresis plot of \(F_{STF}\) versus velocity \(v\) [m/s]. The lines in (c) and (d) are colored by time, from blue (\( t = 25 \) s) to red (\( t = 40 \) s), as shown in the colorbar.}
    \label{fig:Resonance-Constant-15}
\end{figure}

To determine whether the nonlinear and irregular behavior of \(F_{STF}\) persists at a constant excitation frequency, Fig.~\ref{fig:Resonance-Constant-15} examines the system at \(\omega = 15\) rad/s over \( t \in [0, 60] \) seconds. Subfigure (a) shows the time histories of \(x\) and \(F_{STF}\), where their extrema remain correlated, similar to the resonance region with varying frequency. Subfigure (b) confirms that \(\omega\) is constant at 15 rad/s. Subfigures (c) and (d) present the 3D and 2D hysteresis plots of \(F_{STF}\) versus \(v\), respectively, with loop areas indicating substantial energy dissipation. 

In this region, the figures above show that $F_{STF}$ exhibits highly damped and slightly irregular behavior, effectively dissipating energy. The transition from a low-viscosity state to a high-viscosity state may occur in this region, leading to increased damping.

\subsubsection{Post-resonance region}
\begin{figure}
    \centering
    \includegraphics[width=\textwidth]{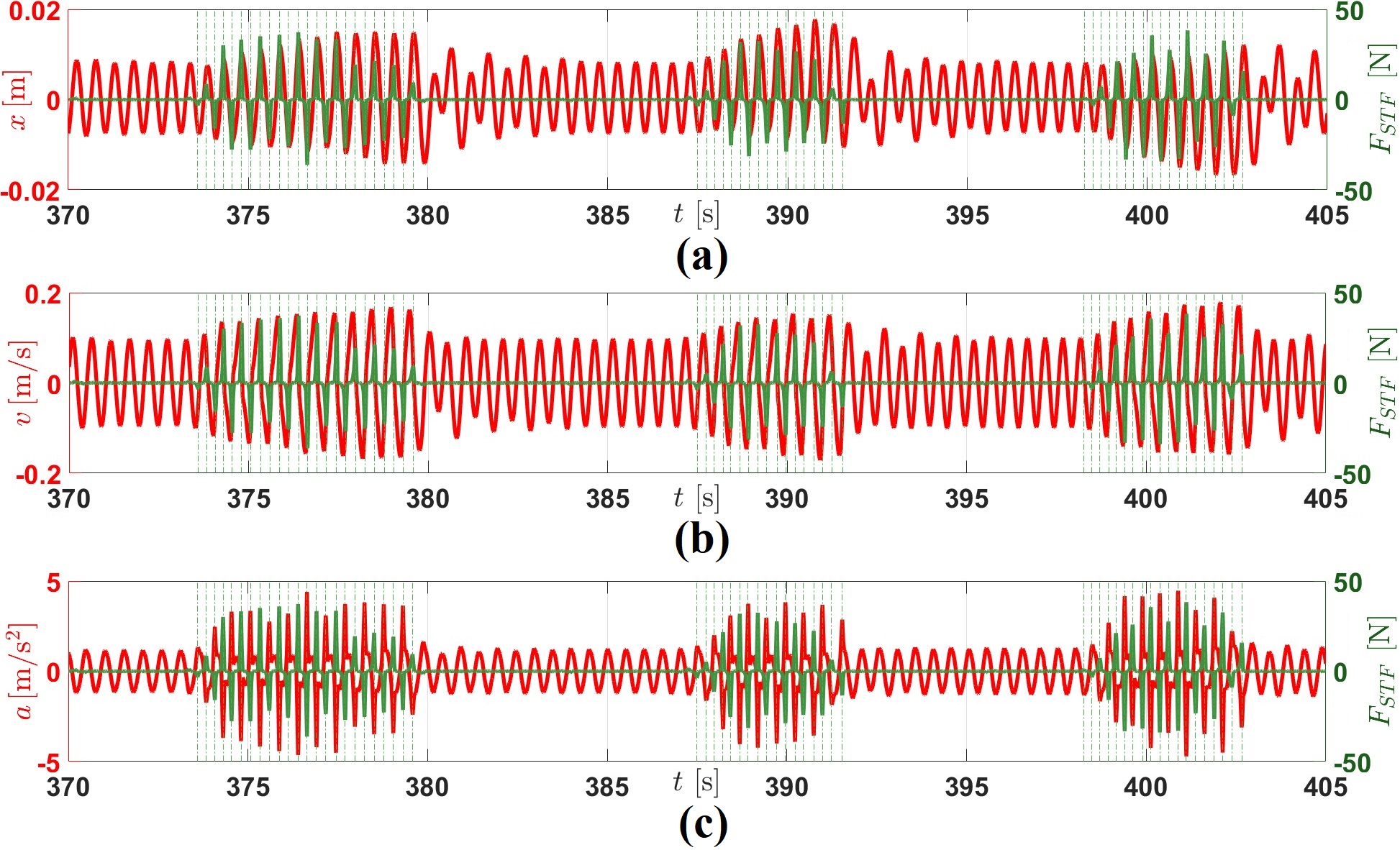}
    \caption{Time histories of (a) displacement \(x\) [m], (b) velocity \(v\) [m/s], and (c) acceleration \(a\) [m/s\(^2\)] (red, left y-axis) compared with the STF force \(F_{STF}\) [N] (green, right y-axis) during the time interval \( t \in [370, 405] \) seconds in the post-resonance region. The excitation frequency \(\omega\) increases at a constant rate of \( 0.0333 \, \text{rad}/\text{s}^2 \). Vertical dashed green lines indicate the extrema of \(F_{STF}\).}
    \label{fig:Post-Resonance-TimeHistory}
\end{figure}

As in the previous subsection, we examine how the STF force (\(F_{STF}\)) in the post-resonance region relates to displacement (\(x\)), velocity (\(v\)), and acceleration (\(a\)), focusing on extrema correlations. The figure below illustrates these relationships over time. From Fig.~\ref{fig:Post-Resonance-TimeHistory}, \(F_{STF}\) exhibits repeated impact-like spikes. Between these spikes, the force drops to zero, illustrating an on-off force pattern. 

Additionally, the peaks of the velocity \(v\) occur near the zero crossings of \(F_{STF}\), albeit with slight delays. The extreme values of \(F_{STF}\), highlighted by vertical dashed green lines, align closely with those of both displacement \(x\) and acceleration \(a\); however, a minor phase lag is observed with \(x\). The correspondence is most precise with \(a\), where the peaks and troughs of \(F_{STF}\) coincide exactly with those of acceleration. This strong correlation suggests that \(F_{STF}\) is directly proportional to \(a\), in line with Newton’s second law, while the small delays in \(x\) and \(v\) reflect the system’s response. 

\begin{figure}
    \centering
    \includegraphics[width=\linewidth]{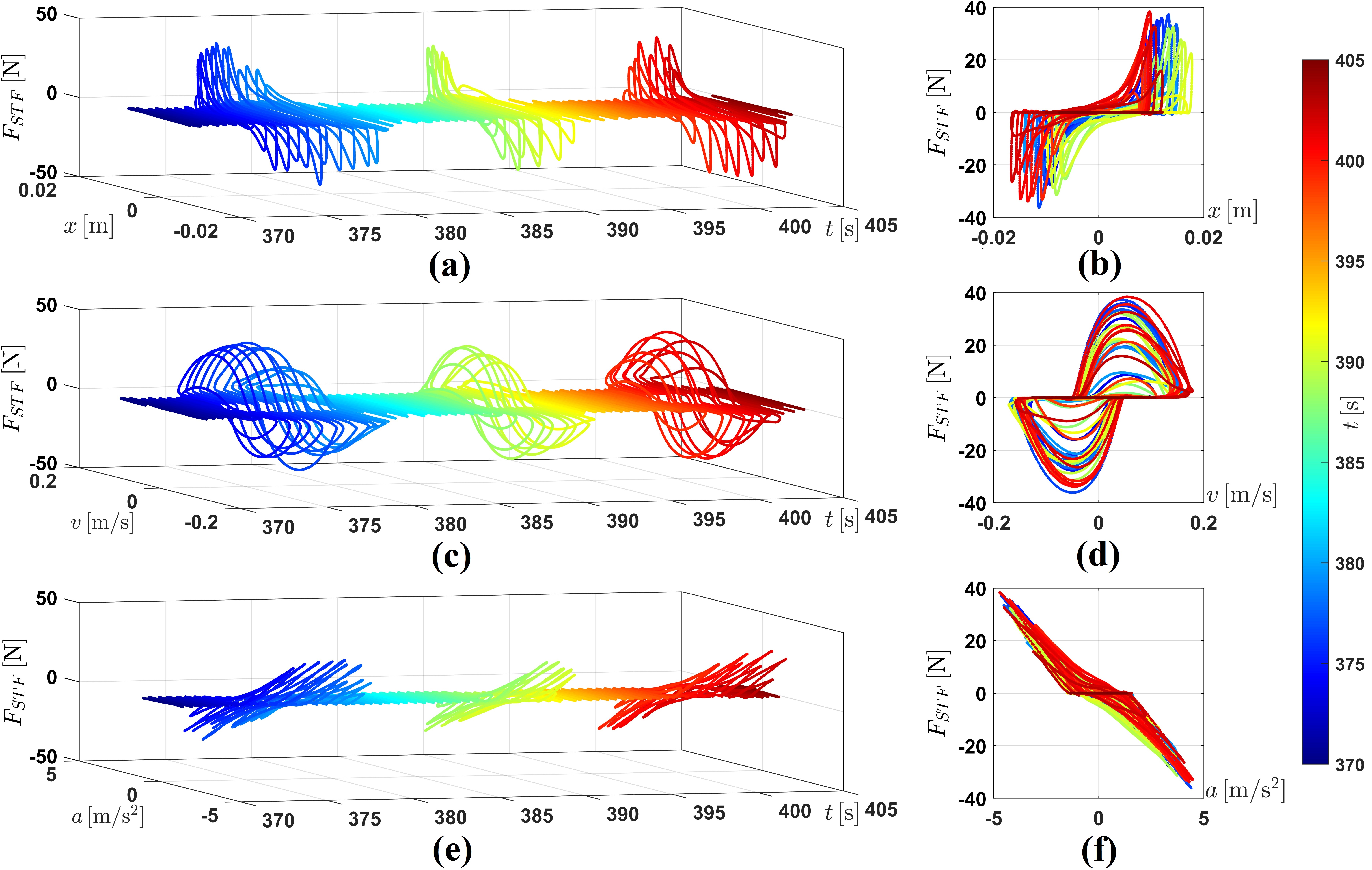}
    \caption{Hysteresis plots of \(F_{STF}\) versus (a) 3D displacement \(x\) [m], (b) 2D displacement \(x\) [m], (c) 3D velocity \(v\) [m/s], (d) 2D velocity \(v\) [m/s], (e) 3D acceleration \(a\) [m/s\(^2\)], and (f) 2D acceleration \(a\) [m/s\(^2\)] during the time interval \( t \in [370, 405] \) seconds in the post-resonance region. The lines are colored by time, from blue (\( t = 370 \) s) to red (\( t = 405 \) s), as indicated by the colorbar.}
    \label{fig:Post-Resonance-hystersis}
\end{figure}

The hysteresis plots in Figure~\ref{fig:Post-Resonance-hystersis} highlight an on-off force pattern and demonstrate the system's nonlinear behavior. In subfigures (a) and (b), the \( F_{STF} \)-\( x \) loops display irregular and spiked enclosed areas, indicative of impact-like spikes. Similarly, the \( F_{STF} \)-\( v \) loops in subfigures (c) and (d) enclose pulsed areas, suggesting repeated impact-like events. In contrast, subfigures (e) and (f) present nearly linear \( F_{STF} \)-\( a \) loops with minimal enclosed areas, reflecting a proportional relationship between force and acceleration. 

Compared to the resonance region, the post-resonance region shows considerably less energy dissipation over time despite the larger loop areas in the \( F_{STF} \)-\( v \) plots, as evidenced by the force being off for certain intervals. 

\begin{figure}
    \centering
    \includegraphics[width=\textwidth]{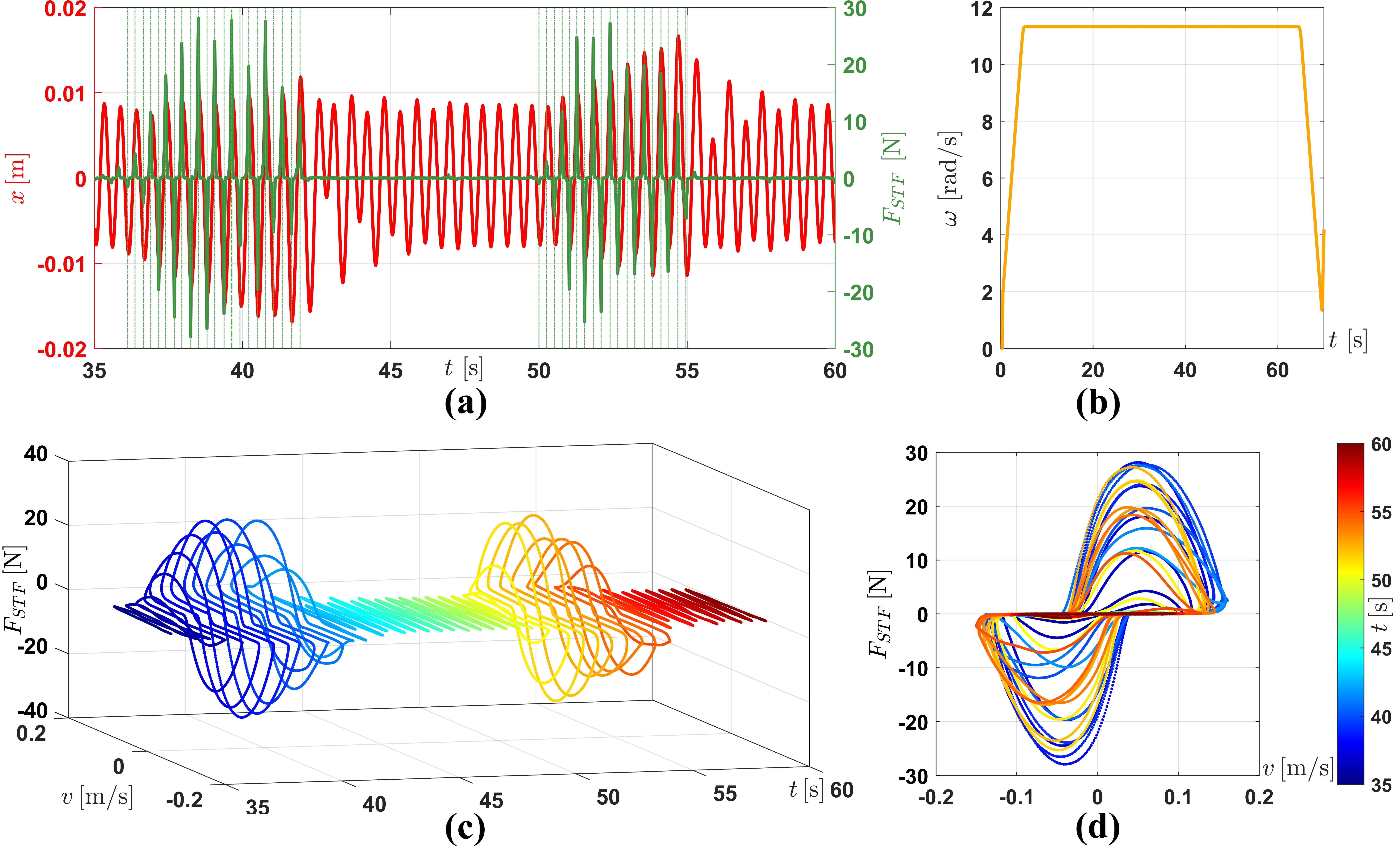}
    \caption{Plots at a constant excitation frequency \(\omega = 11.3\) rad/s: (a) Time histories of displacement \(x\) [m] (red, left y-axis) and STF force \(F_{STF}\) [N] (green, right y-axis), (b) Time history of the excitation frequency \(\omega\) [rad/s] over \( t \in [35, 60] \) seconds, (c) 3D hysteresis plot of \(F_{STF}\) versus velocity \(v\) [m/s], and (d) 2D hysteresis plot of \(F_{STF}\) versus velocity \(v\) [m/s]. The lines in (c) and (d) are colored by time, from blue (\( t = 35 \) s) to red (\( t = 60 \) s), as shown in the colorbar.}
    \label{fig:Post-Resonance-Constant-22}
\end{figure}

Fig.~\ref{fig:Post-Resonance-Constant-22} examines whether these impact-like cycles persist under a fixed excitation frequency in this region. Subfigure (a) confirms a similar phase relationship as in the varying-frequency case. Subfigure (b) shows \(\omega\) remains constant at 22 rad/s. In (c) and (d), the large enclosed areas in the \(F_{STF}\)-\(v\) loops reveal that repeated impact-like events still occur even without changing \(\omega\).

In this region, \(F_{STF}\) exhibits impact-like spikes, dropping to zero between them and forming an on–off force pattern. A possible physical explanation is that the STF undergoes a dynamic jamming transition, as described by Waitukaitis \& Jaeger \cite{Waitukaitis2012}, in which part of the fluid solidifies into a plug that moves with the plate. As this jammed region grows, it may act as an added mass and intermittently contact the container walls, producing the sharp impact-like force peaks observed in the experiments. Other potential mechanisms—such as cavitation, flow separation, vortex shedding, or surface waves—could also contribute to the irregular force behavior; however, these effects typically generate much smaller forces and are therefore considered less likely as primary causes. At present, these interpretations remain probabilistic hypotheses. Further evidence, such as high-speed imaging or local rheological measurements, would be required to confirm the exact mechanism.

\subsection{STF impact: frequency domain comparison with and without STF} \label{STF Impact: Frequency Domain Comparison with and without STF}
Frequency domain analysis provides a clear comparison to evaluate the impact of STF on oscillator dynamics.
\begin{figure}
    \centering
    \includegraphics[width=\textwidth]{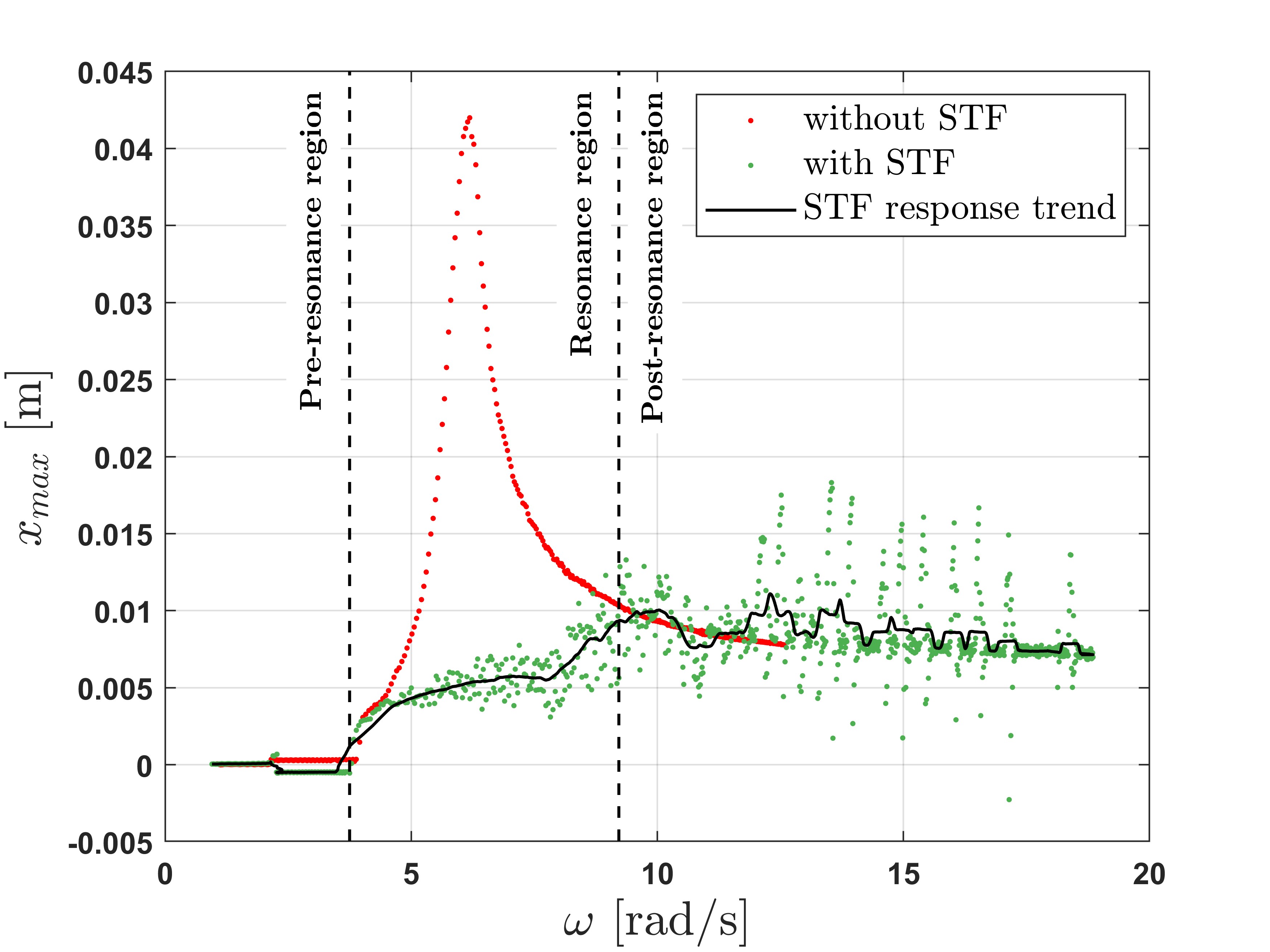} 
    \caption{Frequency response analysis comparing system behavior without STF (red dots), with STF (green dots), and the averaged STF response trend (black line).}
    \label{fig:WithAndWithoutSTF}
\end{figure}
In Fig.~\ref{fig:WithAndWithoutSTF}, red dots represent experimental data for the system without STF (empty container), while green dots correspond to the system with STF (the container filled with STF), as shown in Fig.~\ref{fig:ExperimentalSetup}. To better interpret the overall influence of STF, an averaged trend of the green dots, representing the overall behavior of the system with STF, is plotted as a black line to reduce irregular variations. The black line is obtained through:

\begin{enumerate}
    \item Identifying significant variations using a strict threshold and replacing outliers with the average of adjacent points.
    \item Applying a moving average filter to smooth fluctuations and create a clearer trend.
\end{enumerate}

As observed in the figure, a significant resonance peak occurs at approximately 6.2 rad/s, with a maximum amplitude of 0.042 m, indicating the system’s natural frequency. The response can be divided into three regions:

\begin{itemize}
    \item \textbf{Pre-resonance region (0–3.75 rad/s):} The system demonstrates low-amplitude, stable behavior with no signs of irregular motion and no noticeable influence from STF. This suggests that STF does not interfere with the system at lower frequencies.

    \item \textbf{Resonance region (3.75–9.22 rad/s):} The system enters an irregular state at approximately 4.5 rad/s. STF significantly reduces resonance amplitudes, substantially decreasing the peak amplitude, effectively suppressing unwanted vibrations. The irregular behavior begins at the onset of this region.
    
    \item \textbf{Post-resonance region (\(> 9.22 \, \text{rad/s}\)):} The system stabilizes at a lower amplitude of approximately 0.008 meters, with some fluctuations and irregular behavior. The black line (STF response trend) shows an initial increase in fluctuations, representing irregular behavior, which then decreases as the frequency increases. While the STF introduces slight enhancements in the high-frequency response at a few points, accompanied by increased irregular behavior in this region, the corresponding response levels are still significantly lower than the peak amplitudes observed in the system without STF.
\end{itemize}
After the resonance region, with further increases in excitation frequency, the irregular behavior subsides, and the system’s frequency response with STF more closely resembles that of the system without STF.  Therefore, a similar pattern might be expected for the second resonance frequency, meaning STF has minimal influence before reaching this frequency. However, since the second resonance frequency (without STF) was not studied here, this cannot be confirmed with complete certainty.

\section{Conclusion} \label{Conclusion}

In this study, we explored the response of shear-thickening fluids (STFs) under harmonic excitation to understand their influence on oscillating systems. Using an experimental setup with an unbalanced rotor and a vibrating plate submerged in an STF-filled container, we characterized STF behavior across three frequency regions: pre-resonance, resonance, and post-resonance.

In the pre-resonance region, the STF force was negligible ($F_{STF} \leq 0.2$ N), indicating minimal impact on system dynamics at low frequencies. In the resonance region, the STF acted as an effective damping force, substantially reducing peak vibration amplitude. This significant attenuation highlights the STF's ability to suppress resonance-induced vibrations, which is crucial for stabilizing oscillating systems. In the post-resonance region, the STF exhibited an impact-like, on-off force pattern, introducing irregular behavior with repeated spikes and zero-force intervals. Despite this, vibration amplitudes stabilized at approximately 0.008 meters, lower than without STF.

Overall, STFs enhanced vibration control across the frequency spectrum, particularly in the resonance region. While some irregular behavior occurred during and after resonance, it did not diminish the STF's effectiveness. This study advances our understanding of STF dynamics under harmonic excitation and paves the way for integration into engineering applications.

\section*{Author contribution statement}
Mohammad Parsa Rezaei contributed to the conceptualization, investigation, methodology development, validation, software implementation, formal analysis, visualization, original draft preparation, as well as reviewing and editing the manuscript. Grzegorz Kudra was involved in the conceptualization, investigation, methodology, validation, software, formal analysis, writing, review, and editing, and provided supervision throughout the project.
Krzysztof Witkowski contributed to supervision, investigation, software development, and formal analysis. Grzegorz Wasilewski participated in the investigation, software development, and formal analysis.
Jan Awrejcewicz contributed to supervision, investigation, and formal analysis.

\section*{Acknowledgements}
This study has been supported by the Polish National Science Center, Poland, under the grant PRELUDIUM 22 No. 2023/49/N/ST8/00823. This article was completed while the first author, Mohammad Parsa Rezaei, was a doctoral candidate at the Interdisciplinary Doctoral School at Lodz University of Technology, Poland. For Open Access, the authors have applied a CC-BY public copyright license to any Author Accepted Manuscript (AAM) version arising from this submission. The author thanks Prof. Steve Shaw (Florida Institute of Technology) for his helpful feedback on the manuscript.

\section*{Declaration of competing interest}
The authors certify that no personal relationships or known competing financial interests could have appeared to influence the research presented in this paper.

\section*{Data availability statement}
The data supporting these findings (raw and processed) are available at: \href{https://repod.icm.edu.pl/dataset.xhtml?persistentId=doi:10.18150/FDFWSQ\&faces-redirect=true}{https://repod.icm.edu.pl}.

\end{document}